\let\orilabel\label % 1. Save the clean kernel label rule

\documentclass[
    aps,
    prb,
    reprint,
    superscriptaddress,
    nofootinbib,
    floatfix
]{revtex4-2}
\let\label\orilabel % 2. Restore it to overwrite revtex's change
        
\usepackage{amsmath,amssymb,amsfonts}
\usepackage{upgreek} % upright Greek (e.g. \upmu for electrochemical potentials)
\usepackage{graphicx}
\usepackage{bm}
\usepackage{physics}
\usepackage{xcolor}
\usepackage{booktabs}
\usepackage{microtype}
\usepackage[T1]{fontenc}
\usepackage[scaled]{helvet} % Helvetica (Arial substitute) for \sffamily text
\usepackage{tikz}
\usetikzlibrary{shapes.geometric, arrows.meta, positioning, fit, backgrounds}
\usepackage{pgfplots}
\pgfplotsset{compat=1.18}
\usepackage{sansmath} % sans-serif digits in math-mode tick labels
\usepackage[english]{babel}

\usepackage{natbib} 
\tikzset{
  block/.style  = {rectangle, rounded corners=4pt, draw=black!70,
                   fill=#1, text width=5.2cm, align=center,
                   minimum height=0.75cm, font=\small},
  block/.default = blue!12,
  arrow/.style  = {-{Stealth[length=5pt]}, thick},
  label/.style  = {font=\footnotesize\itshape, text=black!60},
}
\usepackage[colorlinks=true,citecolor=blue,linkcolor=blue,urlcolor=blue]{hyperref}
\usepackage{xurl} % allow long URLs to break across lines
\usepackage{cleveref} 
  
\begin{document}
               
% --- Title ---
\title{Machine Learning for Charge State Characterization of Isolated Double Quantum Dots}
      
% --- Authors & affiliations ---
\author{Hyma H. Vallabhapurapu}
\email{h.vallabhapurapu@diraq.com}
\affiliation{Diraq, Sydney, New South Wales, Australia}
  
\author{Marco Candido}
\affiliation{Diraq, Sydney, New South Wales, Australia}
\affiliation{School of Electrical Engineering and Telecommunications, University of New South Wales, Sydney, New South Wales, Australia}  
  
\author{Krishna Choudhary}
\affiliation{Diraq, Sydney, New South Wales, Australia}
  
\author{Paul Steinacker}
\affiliation{Diraq, Sydney, New South Wales, Australia}
 
\author{Ensar Vahapoglu}
\affiliation{Diraq, Sydney, New South Wales, Australia}

\author{Chris Escott}
\affiliation{Diraq, Sydney, New South Wales, Australia}
    
\author{Wee Han Lim}
\affiliation{Diraq, Sydney, New South Wales, Australia}
\affiliation{School of Electrical Engineering and Telecommunications, University of New South Wales, Sydney, New South Wales, Australia}

\author{Andre Saraiva}
\affiliation{Diraq, Sydney, New South Wales, Australia}
\affiliation{School of Electrical Engineering and Telecommunications, University of New South Wales, Sydney, New South Wales, Australia}

\author{Nard Dumoulin Stuyck}
\affiliation{Diraq, Sydney, New South Wales, Australia}
\affiliation{School of Electrical Engineering and Telecommunications, University of New South Wales, Sydney, New South Wales, Australia}

\author{MengKe Feng}
\affiliation{Diraq, Sydney, New South Wales, Australia}
               
% --- Date ---
\date{\today}
                 
% --- Abstract ---
\begin{abstract}

Scaling semiconductor quantum dot arrays toward fault-tolerant quantum
computation demands efficient tuneup of spin qubits, a task that hinges
on the analysis of charge stability maps (CSMs) and is still largely
performed by hand.  While machine learning has been applied
extensively to this analysis in the conventional reservoir-coupled
regime, the growing use of isolated-mode operation across
gate-defined quantum dot platforms opens a
distinct setting for automated tuning that has so far received only
limited attention.  Isolated-mode CSMs, whose charge transitions appear
as near-vertical lines, are well suited to compact, task-specific
models; we find that convolutional neural networks with under a million
parameters analyse these images accurately.  We present two such
models, trained and evaluated on CSMs from 32 silicon
metal-oxide-semiconductor (SiMOS) double-quantum-dot devices measured at
${\sim}1$\,K with an automated cryogenic probing system, with 16 devices
used for training and 16 held out and scored against hand-labelled
ground truth to assess cross-device generalization.  \textsc{CSMClassifier} screens each image for charge instability and sensor artefacts, achieving 94\,\% macro-averaged
accuracy across three quality categories on a 2{,}407-image held-out set.
\textsc{ChargeLineNet} localises charge-transition lines and reads off the
electron occupancy, reaching 95.3\,\% exact line-count accuracy on a 1{,}131-image held-out set.  Combined into a single
pipeline, in which the classifier first screens each image and passes only
those judged clean to the line counter, the two models together return the
correct electron occupancy for 93.8\,\% of clean held-out images.  Pre-training on
order-$10^5$ synthetic images makes this
accuracy strikingly label-efficient: fine-tuning the synthetic model on
hand-labelled experimental data matches training from scratch when
labels are plentiful, but retains above 90\,\% accuracy when labels are
scarce, where the from-scratch model collapses.  Together the two models
occupy just 6.5\,MB and run under 60 ms per image on standard laboratory hardware, a key step
towards scalable, automated characterization of quantum-dot devices.
\end{abstract}

\maketitle
                     
% =====================================================
\section{Introduction}
\label{sec:introduction}

% ---- Double quantum dot device overview (4-panel) ----
\begin{figure*}[t]
\centering
\includegraphics{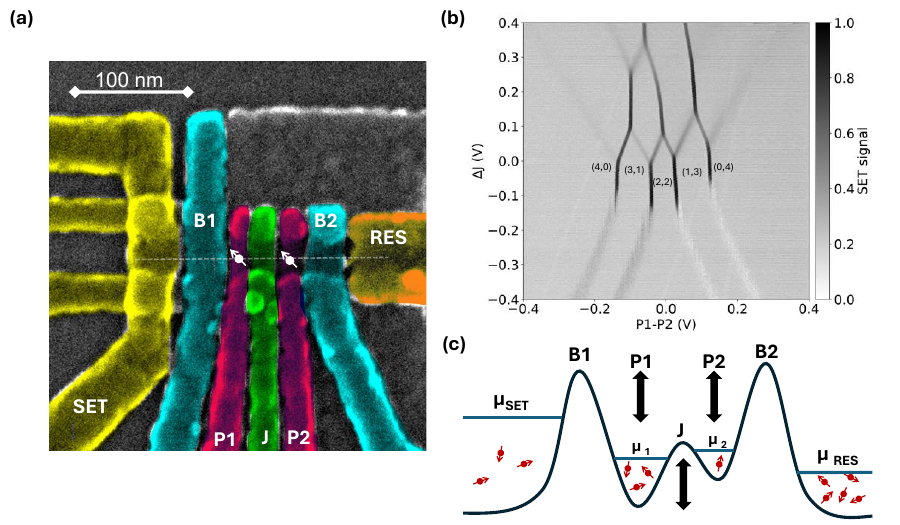}
\caption{Overview of the double-quantum-dot (DQD) device and its charge
stability map.
(a)~Representative Scanning-electron-microscope (SEM) image of a DQD
device with an integrated single-electron-transistor (SET) charge sensor,
with the barrier gates B1 and B2 controlling the tunnel barriers to the
SET and to the reservoir respectively.  The image is of a device fabricated
in an academic cleanroom, shown for illustration of the gate layout only;
all charge-stability data in this work are from imec-fabricated devices of
nominally identical architecture.
(b)~A measured charge stability map (CSM): the sensor current recorded
while sweeping two plunger-gate voltages (P1-P2), with charge-transition lines
marking the boundaries at which the electron occupancy changes.  Because
the device is operated in isolated mode, the total electron number is
fixed (here $N=4$), so each region is labelled $(N_1, N_2)$ by the number
of electrons on the two dots and every charge-transition line marks the
transfer of a single electron from one dot to the other, taking the
configuration through $(4,0), (3,1), \dots, (0,4)$ as the plunger voltages
redistribute the fixed charge between the dots.
(c)~Illustration of the electrostatic potential wells and the gates used
for the voltage sweeps (taken along the dotted line cross section), indicating how the plunger gates (P1, P2 and J) set the electrochemical
potentials of each dot ($\upmu_{1}$ and $\upmu_{2})$.}%
\label{fig:device_overview}
\end{figure*}

Spin qubits in gate-defined quantum dot devices are a promising platform for scalable computing ~\cite{Loss1998-xx,elzerman2004single,petta2005coherent,angus2007gate}. Across material systems, realizing this promise requires bringing large numbers of near-identical devices into operation, and this begins with the same electrostatic tuning problem: shaping the confinement potentials and verifying the charge occupancy of each dot. Silicon metal-oxide-semiconductor (SiMOS) devices are a particularly prominent example, having demonstrated operation at higher temperatures \cite{yang2020operation, petit2020universal} and, importantly, compatibility with established complementary-MOS (CMOS) processes \cite{dumoulin2026cmos} that enable tight integration with the classical control hardware required for qubit operation~\cite{bartee2025spin, Members_of_the_HRL_Quantum_Team2026-vp}. Recently, SiMOS double quantum dot devices fabricated on industrial foundry lines have demonstrated high-fidelity two-qubit operation \cite{steinacker2025industry}, and this has been extended to the coherent operation of an eight-dot array configured as multiple double-dot pairs \cite{nickl_eight-qubit_2026}, with both Steinacker et al.\ and Nickl et al.\ reporting the successful tuning and operation of multiple quantum dots exhibiting high-quality qubit characteristics. While this marks an important step towards reproducible, high-volume qubit fabrication, manually bringing each such device into operation remains a significant bottleneck common to all gate-defined quantum dot platforms. Although the devices we study here are SiMOS double quantum dots (DQDs), the charge-analysis problem we address depends only on the charge-stability-map phenomenology of a gate-defined double dot operated in isolated mode; the approach is therefore in principle applicable to any such DQD platform regardless of host material, and we demonstrate it here on SiMOS devices.

One of the first challenges (among many \cite{Zwolak2024-tf,zwolak2023colloquium}) in tuning such quantum dot devices is reliably loading the exact number of electrons into the dots, which requires careful shaping of the dot potentials and accurately counting the electrons actually loaded. This counting problem is not unique to two-dot devices: even in larger multidot devices ($>2$ dots), the charge stability maps used for tuning are still acquired by sweeping two plunger gates at a time, so each measurement reduces to a DQD problem~\cite{nickl_eight-qubit_2026}, making DQD data an excellent basis for training models that generalize to the pairwise measurements encountered in scaling up to larger arrays. This problem of making sense of electron occupancy has been conventionally approached in the literature as an image analysis problem employing vision-based machine learning (ML) models \cite{Members_of_the_HRL_Quantum_Team2026-vp, samaha2026automatic, muto2026automatic,kalantre2019machine,durrer2020automated,zwolak2020autotuning,schuff2026fully,Hader2025-ew,Darulova2020-ba,yon2025experimental,losert2025automated,diazmoreno2025benchmarking}. These models have been developed predominantly for the conventional reservoir-coupled configuration, in which the dots remain tunnel-coupled to a nearby charge reservoir; as we discuss below, the isolated-mode regime that is our focus here presents a distinct analysis problem that this body of work does not address.

The device that produces these charge stability maps (CSMs) is shown in \cref{fig:device_overview}(a), a top-down scanning electron microscopy image of a SiMOS DQD device. The single electron transistor (SET) is used as an integrated charge sensor of the device, decoupled from the quantum dots by a tuneable barrier (B1). The electrons are loaded into the double dots formed under plunger gates P1 and P2 via a reservoir that is coupled via another tuneable barrier (B2). The quantum dot potentials themselves are separated from each other by the tuneable barrier (J). The barriers and plungers together set the electrochemical potentials $\upmu$ of the dots relative to each other, the reservoir and the SET. Charge movements across the dots (triggered by gate voltage sweeps) manifest as sharp changes in the SET current. The resulting changes in electron occupancy of the dots can be clearly visualized as the change in SET signal as a function of the plungers (P1 and P2 swept antisymmetrically) and exchange (J) gate voltages in \cref{fig:device_overview}(b).

An increasingly adopted alternative is to operate such devices in so-called isolated mode \cite{Johnson2005-sp, Bertrand2015-po,yang2020operation}, wherein the barriers (B1 and B2) to the SET and the reservoir are raised after flooding the dots in the low-electron regime, thereby `pinching-off' both the reservoir and the SET, which can itself act as a charge reservoir (\cref{fig:device_overview}(c)). Beyond simplifying operation, isolating the dots from the reservoir restores the full tunability of the double-dot system~\cite{Bertrand2015-po}: with the total electron number held fixed, the interdot tunnel coupling can be tuned over several orders of magnitude, allowing the exchange interaction to be controlled and maintained at a charge-noise sweet spot that preserves spin coherence during coherent exchange operations. Isolation additionally suppresses parasitic reservoir-induced effects, such as photon-assisted tunneling, during spin manipulation~\cite{Bertrand2015-po}. Charge sensing of such an electrically isolated double dot was first demonstrated by Johnson et al.~\cite{Johnson2005-sp}, who showed that once the tunnel coupling to the leads is suppressed the total electron number is latched and the familiar honeycomb stability pattern collapses into a simpler set of transitions reflecting only the redistribution of that fixed charge between the two dots.

This structural simplification makes the configuration well suited to automated image analysis: charge occupancy can be determined by counting long near-vertical lines that characteristically fade into the background due to reduced tunnel-coupling as a function of the exchange gate barrier (J) \cite{candido2025investigation}. Such charge stability maps lend themselves to simpler analysis, as long as the vertical lines of interest are accurately identified and located. This apparent simplicity might suggest that a classical line-counting heuristic (for instance a Hough transform or a column-wise projection followed by peak detection) would suffice. In practice such approaches are prone to miscounting: the near-vertical plunger-gate transitions are accompanied by branching structures arising from electrons loading into parasitic dots formed under the device gates, and by background transitions and sensor artefacts, all of which register as additional edges or peaks that a geometry-agnostic counter cannot distinguish from the genuine plunger-gate lines. Reliably counting only the transitions of interest therefore requires a model that has learned to recognise the plunger-gate lines and suppress the spurious ones. This is the learned, orientation-selective approach we adopt here. We quantify this in \cref{app:classical}: a carefully tuned classical counter, optimised on a subsample of the training data (never the held-out devices), reaches only $61\,\%$ exact line-count accuracy, well below the model developed here, and this figure is insensitive to how its hyperparameters are tuned.

Isolated mode is not without drawbacks, however: unless the confinement barriers, together with the barriers to the SET and the reservoir (B1 and B2), are carefully tuned, spurious tunneling events still allow electrons to jump into and out of the dots within a single scan, rendering the charge state unstable, and the SET sensor itself may drift or be disturbed and require retuning. Reliably detecting such unstable or disturbed images is therefore an essential first step, and calls for a CSM quality-classification tool. The value of screening measurements for quality before analysis has been recognised in the reservoir-coupled setting, where a data-quality-control module acts as a `gatekeeper' that admits only reliable data to a downstream state classifier~\cite{ziegler2022robust, Darulova2020-ba}. We take inspiration from this and bring an analogous quality-screening step to the isolated-mode regime, with categories tailored to its characteristic failure modes. Such a tool does more than reject uninterpretable images: by flagging any departure from a well-resolved image (charge instability, sensor drift, or obscuring background structure), it prompts the operator to retune the barriers or SET sensor before acquiring further data, rather than waiting passively for interpretable images to appear.

We took an initial step toward isolated-mode analysis in earlier work~\cite{candido2025investigation}, using a U-Net segmentation model to highlight regions of changing tunnel coupling in isolated-mode CSMs as part of an automated cryogenic device-uniformity study. Here we go beyond region segmentation to the two tasks that a tuning pipeline actually requires: per-image quality screening and explicit charge-transition line counting for electron-occupancy readout. This motivates the dedicated models tailored to analyzing charge stability maps acquired in isolated mode that we develop below.

In this work, we implement two convolutional neural network (CNN) based ML models to address the need for analyzing CSM images taken from devices operated in isolated mode. Each has under a million parameters and is deliberately compact, so that it can be trained on the scarce hand-labelled data available for a new device generation and deployed on standard laboratory control hardware. The first model, \textsc{CSMClassifier} (\cref{subsec:confusion_classifier}), is a multi-label quality classifier trained entirely on hand-labeled experimental data that independently assesses each image for cleanliness, charge instability, and ambiguity via per-class sigmoid outputs. The second model, \textsc{ChargeLineNet} (\cref{subsec:lineheatmap}), detects and localizes charge-transition lines via a signed heatmap. In addition, we demonstrate how transfer learning ~\cite{Oquab, Czischek2022-qi,darulova2021evaluation}, using a combination of a large amount of computationally cheap synthetic data and a nominal amount of experimental data, can help to quickly bridge the domain gap and help in generalization to unseen data (\cref{subsec:finetuning}).

We train and evaluate both models on CSMs from 32 double-quantum-dot devices measured at ${\sim}1$\,K using an automated cryogenic probing system~\cite{candido2025investigation}: 16 devices supply the training data and the remaining 16 are held out entirely to test cross-device generalization. On this held-out data, \textsc{CSMClassifier} attains 94\,\% macro-averaged accuracy across the three quality categories, while \textsc{ChargeLineNet} reaches 95.3\,\% exact line-count accuracy (98.6\,\% within $\pm 1$ count), a roughly 34 percentage-point improvement over the synthetic baseline, and a level of accuracy that fine-tuning retains even when hand-labelled data is scarce, where training from scratch collapses. Combined into a single pipeline, in which the classifier first screens each image and passes only those judged clean to the line counter, the two models together return the correct electron occupancy for 93.8\,\% of clean held-out images (93.9\,\% across all held-out images once the correct rejection of degraded ones is counted). Together the two models occupy just 6.5\,MB and are fast enough to run on standard laboratory hardware. \Cref{sec:model} details the two models, and \cref{sec:results} evaluates their performance on experimental data.

% =====================================================
\section{Model and Methods}
\label{sec:model}

Automated analysis of charge stability maps (CSMs) from DQD devices
operated in the isolated-mode regime involves two tasks that are solved
sequentially.  First, each CSM image must be screened for quality: images
exhibiting charge instability, sensor artefacts, or ambiguous features
should be flagged and excluded before quantitative analysis.  Second,
images that pass the quality screen must be analysed to detect and
localise charge-transition lines, whose count directly encodes the
electron occupancy of the dot.

In both cases the input is a single-channel 2D image (sensor current as
a function of two gate voltages) and the model must generalise to
experimentally acquired images that differ in global contrast, noise
level, and background structure.  Both models are implemented in Python
using PyTorch.  All training and inference in this work are performed
offline on pre-recorded CSMs; the models are trained on a single GPU and
are not deployed in real time on a live device.  They share a common
training recipe (the AdamW optimiser, a cosine
learning-rate schedule with a
short linear warm-up, gradient-norm clipping, and the same on-the-fly
augmentation suite of random flips and resolution scaling) and differ
only in the task-specific choices described below and the hyperparameter
values listed in \cref{app:hyperparams}.  Beyond this shared
recipe, the two models use different CNN architectures and different
training strategies, reflecting the different nature of the two tasks;
\cref{subsec:model_comparison} gives a side-by-side comparison of
their shared and distinct design elements.

The experimental data used throughout this work were collected from 32
double-quantum-dot devices, each with an integrated single-electron-transistor
(SET) charge sensor, measured at ${\sim}1$\,K using an automated cryogenic
probing system~\cite{candido2025investigation}.  Importantly, the 32 devices
are not nominally identical: they span a range of designs, differing, for
example, in gate pitch, gate widths, and gate-oxide thickness, and, on top of
these deliberate design differences, each individual device carries its own
idiosyncrasies arising from fabrication imperfections, so that no two devices
produce identical CSMs.  This design and fabrication spread makes
holding out entire devices a stringent test of cross-device generalisation:
the first 16 devices supply the training data (with a random 95/5
train/test split within those devices) and the remaining 16 are held out
entirely as the validation set used to assess cross-device
generalisation (\cref{sec:results}).  The in-device test split is kept
small deliberately: hand-labelled data is scarce and is best spent on
training, with true generalisation instead measured on the held-out
devices.

The two models are evaluated on the same held-out pool but on the images
relevant to each task.  \textsc{CSMClassifier} is evaluated on all
2{,}407 held-out images that carry a quality label.  \textsc{ChargeLineNet}
is evaluated on the 1{,}131-image subset of these that a human could
reliably count and that therefore received hand-labelled line coordinates
(restricted to the 0--7-line occupancy range on which the model is
evaluated); the remaining images (those with no reliable line count, or
blank or errored acquisitions) are excluded from the line-count
evaluation because they carry no ground-truth count to score against.
Both validation sets thus derive from the same 16 held-out devices, the
1{,}131-image line-detection set being the human-countable subset of the
2{,}407-image quality-screening set.  \Cref{tab:splits} in
\cref{app:splits} consolidates all image counts and device splits.

\textsc{CSMClassifier} is a multi-label quality classifier that
independently assesses each CSM image for three properties (clean,
unstable, and unclear), applied strictly so that any degraded image is
flagged for corrective action (\cref{subsec:confusion_classifier}).
\textsc{ChargeLineNet}, by contrast, is trained and evaluated on the
broader human-countable population, namely every image to which a
reliable line count can be assigned, including mildly-degraded ones, so
that it counts robustly on the degraded images characteristic of real
measurements.  It outputs a dense signed heatmap from which line
positions and electron occupancy are decoded, and is pre-trained on
synthetic data and fine-tuned on hand-labelled experimental images.

The two models are trained separately, but in deployment they are chained
into a single two-stage pipeline that maps a raw CSM to an electron count.  \textsc{CSMClassifier} runs first as a
quality gate: only images it labels clean are forwarded to
\textsc{ChargeLineNet} (the fine-tuned variant), which then decodes the
line count, while images
flagged as unstable or unclear are rejected and returned to the operator
for re-measurement rather than counted.  We evaluate this combined
pipeline on the held-out images common to both models'
validation sets (2{,}407 images, of which 693 are clean with a reliable
ground-truth count), passing each image through the fixed classifier gate
and then the line counter under five independent heatmap-decoding
configurations, and report the pipeline metrics as the mean and standard
deviation across those five runs (\cref{sec:results}).

% ---- CSMClassifier real-data architecture illustration ----
\begin{figure*}[t]
\centering
\includegraphics[width=\textwidth]{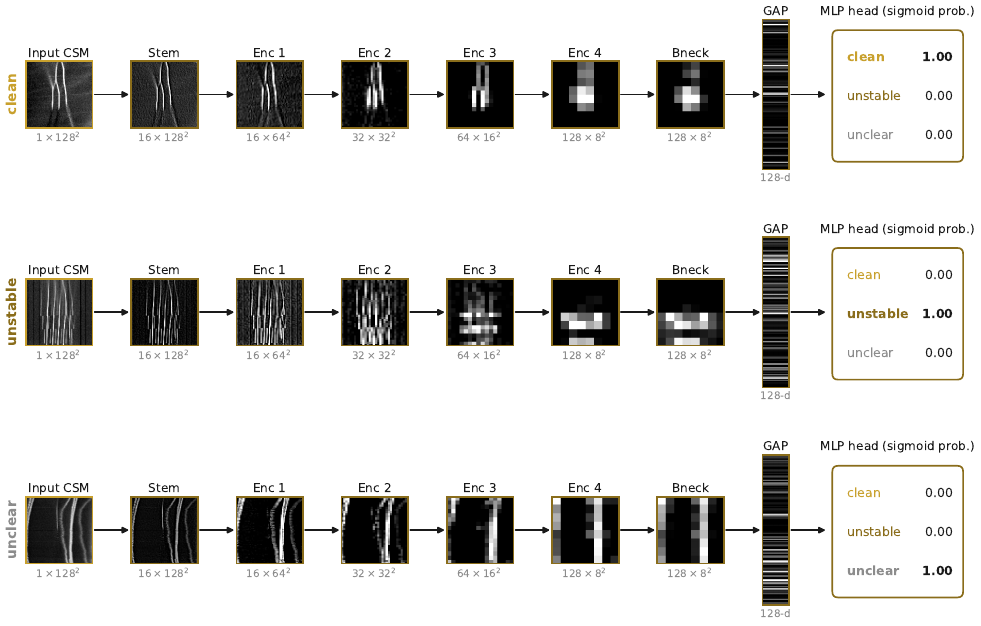}
\caption{\textsc{CSMClassifier} architecture visualised on real CSM
images, one row per quality class (clean, unstable,
unclear).  Each row passes a hand-labelled experimental CSM of
that class through the trained network and shows the actual intermediate
feature maps (only the single highest-variance, i.e.\ most active, channel
is shown at each stage; the remaining channels are omitted to avoid
clutter): input $\to$
histogram-invariant stem (LCN\,$\parallel$\,InstanceNorm) $\to$ four
isotropic encoder stages $\to$ bottleneck $\to$ global average pooling
(GAP) to a 128-dimensional vector.  Spatial and channel dimensions are
annotated below each panel.  The two-layer MLP head then maps the pooled
vector to three independent sigmoid probabilities that do not sum to one,
allowing more than one label to be active at once (e.g.\ simultaneously
unstable and unclear); the predicted (active) class is shown
in bold.  Visually distinct inputs produce distinct
activations, yet each is classified correctly.  The layer-level block
diagram is given in \cref{app:architecture},
\cref{fig:classifier_flow}.}%
\label{fig:classifier_arch}
\end{figure*}

% ---- ChargeLineNet real-data architecture illustration ----
\begin{figure*}[t]
\centering
\includegraphics[width=\textwidth]{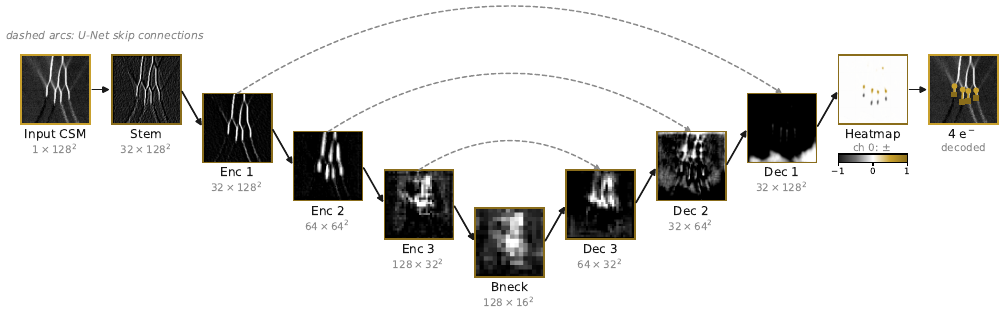}
\caption{\textsc{ChargeLineNet} architecture visualised on a real CSM
image.  A hand-labelled experimental CSM is passed through the trained
network and the actual intermediate feature maps are shown left to right
(only the single highest-variance, i.e.\ most active, channel is shown at
each stage; the remaining channels are omitted to avoid clutter): input $\to$
histogram-invariant stem (LCN\,$\parallel$\,InstanceNorm) $\to$ three
anisotropic encoder levels $\to$ bottleneck $\to$ three U-Net decoder
levels, whose skip connections are drawn as dashed arcs.  The spatial and
channel dimensions after each stage are annotated below the corresponding
panel.  The final $1{\times}1$ head produces the signed heatmap
(channel~0; positive peaks mark line start points, negative peaks mark
end points) shown in the penultimate panel, from which the
charge-transition lines and electron count are decoded by
connected-component labelling and offset-vector readout (rightmost
panel).  For clarity the head's two offset-vector channels
(channels~1--2), which encode the end-to-start pairing, are not shown;
their role is described in \cref{subsec:lineheatmap} and their
effect is reflected in the decoded lines of the rightmost panel.  The layer-level block diagram is given in
\cref{app:architecture}, \cref{fig:chargelinenet_flow}.}%
\label{fig:chargelinenet_arch}
\end{figure*}

% -----------------------------------------------------
\subsection{CSM Quality Classification: \textsc{CSMClassifier}}
\label{subsec:confusion_classifier}

\paragraph{Task Formulation.}

\begin{figure*}[t]
\centering
\includegraphics[width=0.76\textwidth]{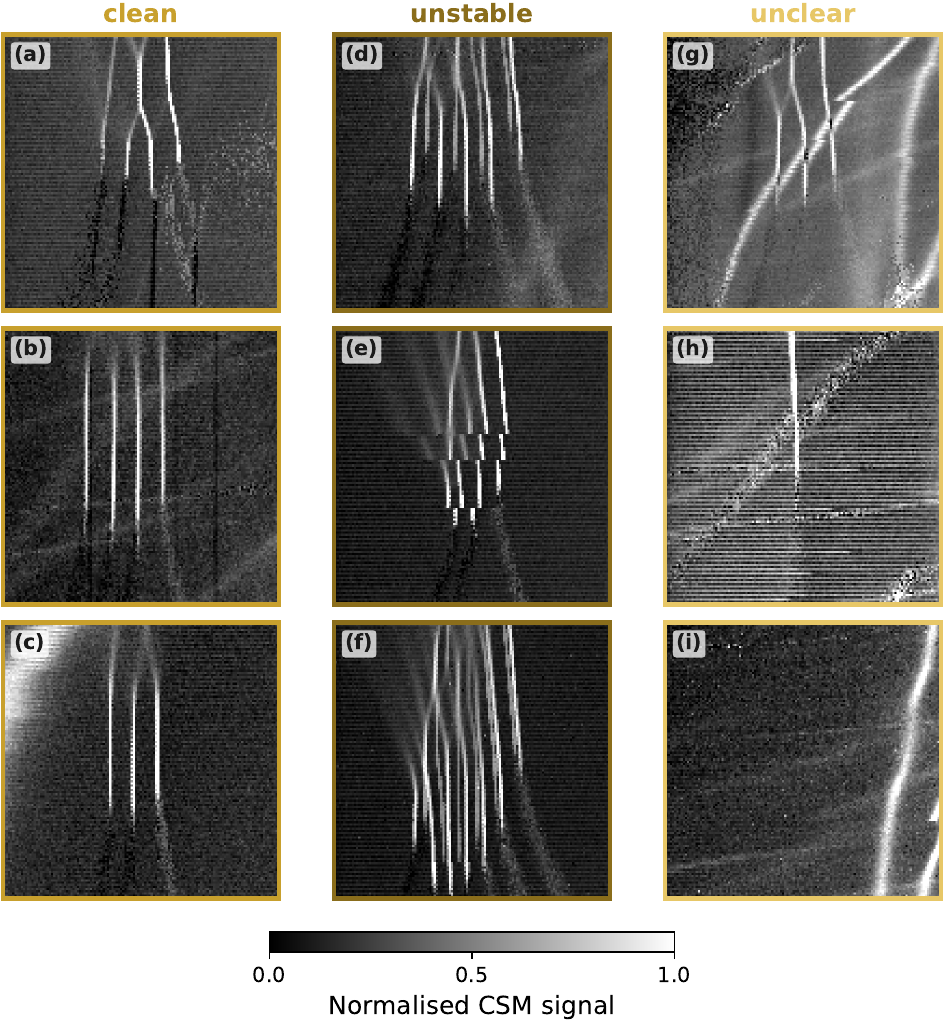}
\caption{Representative CSM images from the three quality classes, one
class per column (three examples each).
\textbf{Left column (a--c), clean:} well-resolved charge-transition
lines with clear contrast against the background, suitable for automated
line detection.
\textbf{Middle column (d--f), unstable:} images in which the electron
occupancy visibly jumps mid-scan, so that transition lines abruptly
appear or disappear part-way across the image (telegraph switching or
spurious tunnelling), leaving no single consistent line count.
\textbf{Right column (g--i), unclear:} images whose line count is
ambiguous even to a human reader, owing to low signal, sensor artefacts,
obscuring background transitions, or SET signal issues, rather than to a
change in the charge state.  An image can be both unstable and unclear at
once.
The \textsc{CSMClassifier} must distinguish these three categories
before line detection is attempted.}%
\label{fig:classifier_examples}
\end{figure*}

Before charge-transition lines can be detected, each CSM image must be
assessed for quality.  Experimental images frequently contain artefacts
that would compromise downstream line detection: charge instability
produces sudden jumps in the sensor signal, SET sensor drift introduces
slow background modulation, and various noise sources can obscure the
charge-transition lines entirely.  The quality classifier assigns each
image independent probabilities for three categories:
\begin{enumerate}
    \item \textbf{Clean}: the image contains well-resolved charge-transition
          lines suitable for automated analysis.
    \item \textbf{Unstable}: the electron occupancy visibly jumps during
          the scan, so that charge-transition lines abruptly appear or
          existing lines abruptly disappear part-way across the image
          (from telegraph switching or spurious tunnelling into and out of
          the dots).  The defining signature is that the number of lines
          is not consistent across the scan, so no single occupancy can be
          read off.
    \item \textbf{Unclear}: the line count is ambiguous to the point that
          even a human could not reliably count the lines, for example
          because of low signal, sensor artefacts, obscuring background
          transitions, or SET signal issues that wash out or overlay the
          charge-transition lines.
\end{enumerate}
The \emph{unstable} and \emph{unclear} labels capture two different
failure modes: \emph{unstable} a genuine physical instability of the
charge state, \emph{unclear} a degradation of image quality that leaves
the count ambiguous to a human reader.  The two are not mutually
exclusive, so the three probabilities are predicted independently via
per-class sigmoid activations rather than as a normalised distribution
that sums to one, allowing an image to be flagged as both unstable and
unclear at once.  The labels are deliberately assigned strictly: any
degraded image is flagged so that the operator can take corrective action
(for example retuning the device or sensor) rather than have a marginal
image passed downstream.  This conservative
screening criterion does not attempt to judge whether the transition
lines of a flagged image could nonetheless still be counted.
Representative examples of each class are shown in
\cref{fig:classifier_examples}.

% ---- Classifier example images ----

\paragraph{Model Architecture.}

The classifier, \textsc{CSMClassifier}, is an encoder-only CNN whose
architecture is illustrated in \cref{fig:classifier_arch}.  The
input is a single-channel CSM image resized to $128 \times 128$ pixels.

The network begins with a histogram-invariant stem that removes global
brightness and contrast differences between devices and cooldowns
(\cref{app:stem}), shared with \textsc{ChargeLineNet} and differing only
in output channel width (\cref{subsec:model_comparison}).

The encoder consists of four stages of isotropic $3 \times 3$ residual
blocks with $2 \times 2$ max-pooling, progressively increasing the
channel width, followed by a bottleneck and adaptive average pooling
that collapses the feature map to a single 128-dimensional vector; a
two-layer multi-layer perceptron (MLP) head maps this vector to three class logits.  Because the
task requires only image-level labels rather than pixel-level
localisation, no decoder or skip connections are needed: the spatial
map is discarded by global average pooling once it has been reduced to
an $8 \times 8$ bottleneck (\cref{fig:classifier_arch}).  Unlike \textsc{ChargeLineNet}, the encoder uses standard isotropic $3
\times 3$ kernels rather than anisotropic multi-branch blocks
(\cref{subsec:model_comparison}): the quality signatures are spatially
diffuse and orientation-agnostic (charge instability manifests as abrupt
horizontal discontinuities, sensor drift as slow vertical gradients, and
ambiguous contrast globally), so no orientation-selective receptive field
is needed.  This is borne out by the
per-class feature maps in \cref{fig:classifier_arch}: the three
quality classes drive visibly different activations that grow
progressively more abstract and spatially coarse through the encoder,
and the class-discriminative signal survives the collapse to a single
pooled vector.  Each output logit
corresponds to one class (clean, unstable,
unclear); at inference, an independent sigmoid is applied to each
logit and a threshold determines the active labels.  Full layer-level
specifications are given in \cref{app:architecture}.

\paragraph{Loss Function.}

The network is trained with binary cross-entropy with logits loss
applied independently to each of the three output classes, with optional
per-class positive weights (computed from the inverse positive-class
frequency) to handle class imbalance.  The full expression is given in
\cref{app:loss_classifier}.

\paragraph{Training Procedure.}

The classifier is trained entirely on hand-labelled experimental data.
No synthetic pre-training is used, because the artefacts the model must
recognise (charge instability, sensor drift, and measurement noise
patterns) are inherently properties of real devices and are difficult to
reproduce faithfully in a simulator.

Beyond the shared training recipe (\cref{sec:model}), per-class
positive weights are enabled to handle class imbalance.  The classifier
is trained on $10{,}060$ hand-labelled CSM images drawn
from 16 devices, spanning all three quality categories; a separate set of
16 held-out devices provides the 2{,}407-image validation set used to
assess cross-device generalisation (\cref{sec:results}).  Note that
\textsc{ChargeLineNet} is fine-tuned on a smaller set of
$6{,}452$ images (\cref{subsec:lineheatmap}): the
human-countable images that carry reliable ground-truth line annotations.

% -----------------------------------------------------
\subsection{Charge-Transition Line Detection: \textsc{ChargeLineNet}}
\label{subsec:lineheatmap}

\paragraph{Task Formulation.}

The central task is to detect and localise charge-transition lines in
CSM images acquired from a DQD device operated in the isolated-mode
regime.  Each CSM image is a two-dimensional map of sensor current as a
function of two gate voltages.  Charge transitions appear as
near-vertical lines whose start (sharp, high-contrast) and end (diffuse,
low-contrast) points encode the electrochemical potential of the dot,
and the number of visible lines directly reports the electron occupancy.
The model is optimised for the low-occupancy regime (fewer
than ${\sim}10$ electrons per double dot), which is the regime of practical
interest: reliable resolution of the $2$-, $4$-, and $6$-electron
configurations enables the search for Pauli spin blockade,
the mechanism used for spin-qubit initialisation and readout \cite{lai2011pauli,seedhouse2021pauli}.

\textsc{ChargeLineNet} operates on the CSM images to which a human can
assign a reliable line count.  This human-countable population is broader
than the strictly clean images: it also includes images with mild
background structure or sensor noise through which the transition lines
can nonetheless be counted.  \textsc{ChargeLineNet} is deliberately
trained and validated on this broader population, rather than on strictly
clean images alone, so that it learns to detect transition lines robustly
in the presence of the mild background structure and sensor noise that
are characteristic of real experimental measurements, rather than only on
idealised images.

A further challenge is that, in many device geometries, additional charge
transitions arise from electrons loading into parasitic dots formed under
the device gates, producing honeycomb-like branching structures in the
CSM.  These transitions do not correspond to the intended plunger-gate
loading and must be excluded from the line count.  As discussed in the
introduction, a geometry-agnostic classical heuristic cannot perform this
selective counting.  The model is instead
trained to detect only the near-vertical charge-transition lines
associated with plunger-gate loading, learning to ignore the branching
features from parasitic device-gate dots through the combination of
anisotropic convolutional kernels (which are selective for near-vertical
orientations) and the training data, which labels only plunger-gate
transitions.  The orientation selectivity of these kernels is
illustrated in \cref{fig:chargelinenet_branches}, which shows how
the six branches of the first encoder block respond to a real CSM: the
tall $7\times1$ and diagonal $3\times3$ branches lock onto the
near-vertical transition lines, while the wide $1\times7$ and
$1\times15$ branches respond to horizontal gate-voltage modulation
instead.

The problem is formulated as signed heatmap plus offset regression,
adapting the keypoint-heatmap paradigm of
object-detection networks~\cite{zhou2019objectsaspoints}
to signed start/end points: the
model outputs a three-channel spatial map $\hat{Y} \in \mathbb{R}^{3
\times H \times W}$.  Channel~0 is a signed heatmap in which positive
Gaussian blobs ($+1$ peak) mark line \emph{start} points and negative
Gaussian blobs ($-1$ peak) mark line \emph{end} points; channels~1--2
encode a per-pixel offset vector $(\Delta x, \Delta y)$ pointing from
each end point towards its paired start point, weighted by the end-point
Gaussian blob so that the offset signal is concentrated at end-point
locations and zero elsewhere.  Ground-truth targets are constructed by
placing isotropic Gaussians of radius $\sigma = 1$\,px at each labelled
start/end coordinate.  This formulation preserves full spatial
information, allows the model to express uncertainty through blob width,
and enables direct end-to-start pairing without a separate matching step:
the offset vectors encode which start belongs to which end, avoiding the
systematic undercounting that a signed heatmap alone suffers when
closely-spaced blobs merge (we detail the end-point anchoring choice and
its robustness in \cref{app:inference}).  A further practical
benefit of this point-based target is that annotation requires only two
clicks per line (the start and end coordinates) rather than a
dense per-pixel segmentation mask, substantially reducing the manual
effort of labelling each image.

\paragraph{Model Architecture.}

\textsc{ChargeLineNet} builds on a standard U-Net encoder--decoder with
skip connections~\cite{ronneberger2015unet}, with task-specific
modifications (anisotropic branches, histogram-invariant stem) tailored to
isolated-mode CSM analysis.  It is a fully-convolutional
encoder--decoder network (\cref{fig:chargelinenet_arch}) whose
design is motivated by three properties of CSM images: (i)~lines span a
wide range of orientations and widths; (ii)~the global intensity
distribution varies substantially between simulator-generated and
experimentally acquired images; and (iii)~start and end points must be
localised to within a few pixels.  These three properties map directly
onto three architectural choices, each visible in
\cref{fig:chargelinenet_arch}: the multi-branch anisotropic
encoder addresses~(i), the shared histogram-invariant stem
addresses~(ii), and the U-Net decoder with skip connections
addresses~(iii).

The network begins with a histogram-invariant stem, shared with
\textsc{CSMClassifier} (\cref{subsec:confusion_classifier}), that
removes global brightness and contrast differences between simulated
training data and experimental test data (\cref{app:stem}).  The encoder then
applies three levels of \emph{anisotropic blocks}, each processing its
input through six parallel convolutional branches spanning a range of
orientations and scales (from a tall $7\times1$ kernel selective for
near-vertical transition lines to a wide $1\times15$ kernel for broad
transitions; \cref{fig:chargelinenet_branches} shows the response
of each branch to a real CSM), halving the spatial resolution and
increasing the channel width at each level.  A U-Net decoder mirrors the encoder, bilinearly
upsampling and concatenating encoder skip connections to restore full
spatial resolution, and a final $1\times1$ convolution produces the
three unbounded output channels (signed heatmap and $(\Delta x, \Delta
y)$ offset vectors).  This restoration is evident in
\cref{fig:chargelinenet_arch}: the feature maps contract to a
coarse $16\times16$ bottleneck and are then progressively resharpened by
the decoder, so that the transition lines re-emerge at full resolution
in the predicted heatmap, giving the spatial precision required by
property~(iii).  The total parameter count is $935{,}283$ for the
default base channel width $c_1=32$.  Full layer-level specifications,
including the anisotropic branch set, dropout rates, and stem constants,
are given in \cref{app:architecture}.

\paragraph{Loss Function.}

The network is trained with a combined endpoint--offset loss that
supervises all three output channels jointly: a foreground-weighted
mean-squared error (MSE)
on the signed heatmap (channel~0), which upweights pixels near blob
centres so that missing a blob costs as much as the background, plus a
masked smooth-$\ell_1$ loss on the offset vectors (channels~1--2) that is
active only at end-point blob locations.  The full expressions are given
in \cref{app:loss_chargelinenet}.

\paragraph{Training Procedure.}

The model is pre-trained on synthetic CSM images and then fine-tuned on
hand-labelled experimental images to close the domain gap; both stages
follow the shared training recipe (\cref{sec:model}).  We describe the
fine-tuning procedure, including the synthetic data generator, the
freeze-mode choice, and the domain-gap augmentation, in
\cref{subsec:finetuning}.

\paragraph{Inference and Line Counting.}

\begin{figure*}[t]
\centering
\includegraphics[width=0.76\textwidth]{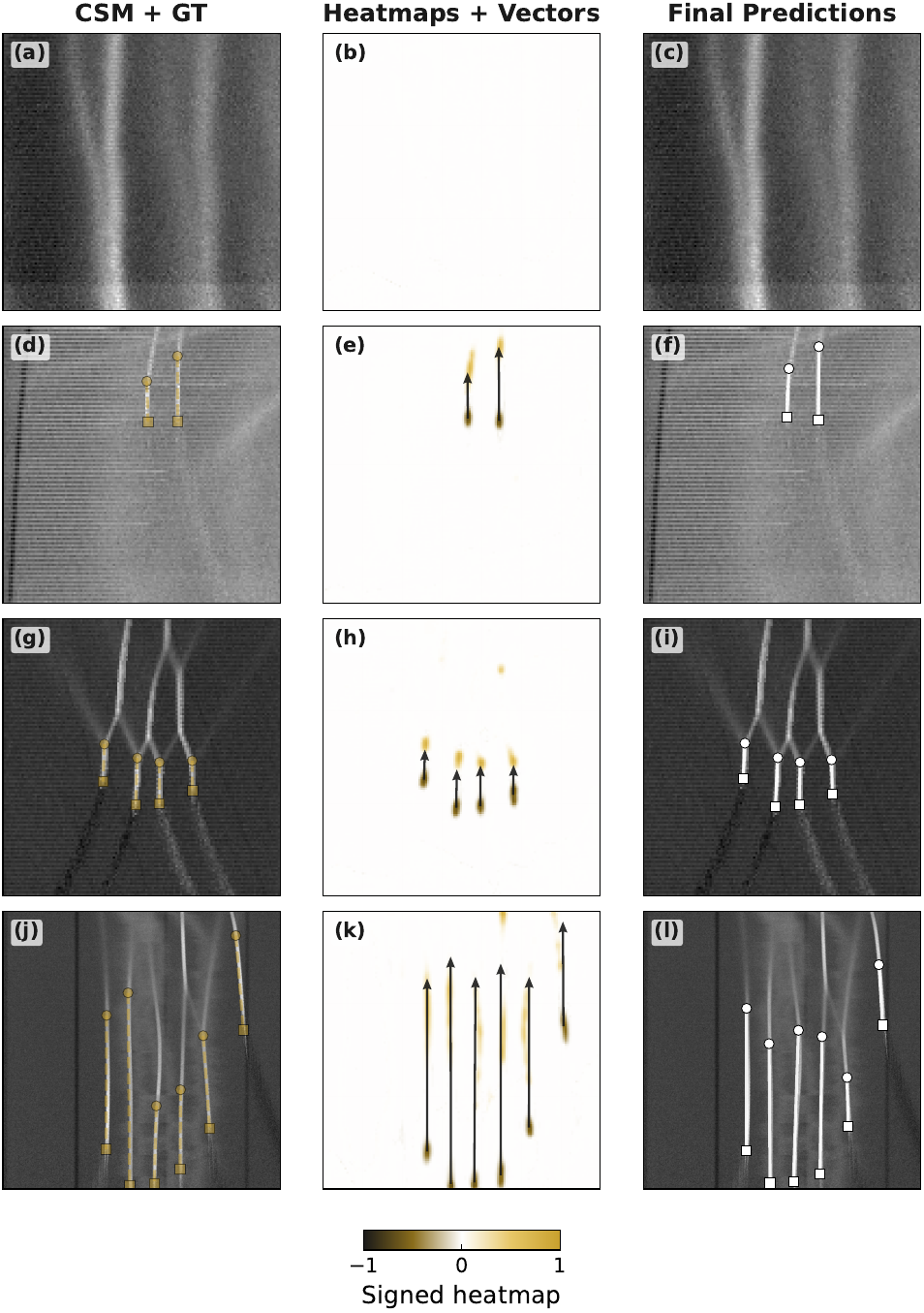}
\caption{\textsc{ChargeLineNet} outputs for three example CSM images
(one per row).
\textbf{Column~1 (a,\,d,\,g,\,j):} preprocessed input CSM.
\textbf{Column~2 (b,\,e,\,h,\,k):} predicted signed heatmap (channel~0), with
positive peaks marking line start points and negative peaks marking line
end points (colour bar).
\textbf{Column~3 (c,\,f,\,i,\,l):} decoded charge-transition lines
overlaid on the input, comparing the ground-truth annotation (GT) with
the model prediction (Pred); markers indicate the detected line
start and end points.
The signed heatmap, together with the predicted offset field (not
shown), provides all information needed to decode line positions and
electron occupancy via connected-component labelling and offset-vector
readout (\cref{subsec:lineheatmap}).}%
\label{fig:chargelinenet_outputs}
\end{figure*}

At inference the three-channel output is decoded into charge-transition
lines (\cref{fig:chargelinenet_outputs} shows the input, predicted
heatmap, and decoded lines for three examples).  End-point blobs are
detected in the signed heatmap, and for each blob the offset vector is
used to locate and pair its start point, which is then snapped onto the
nearest detected start blob so that the vector resolves the pairing
while the blob pins the start to a sharply localised position.  The
electron count is the number of surviving line pairs.  The two
ingredients are complementary and both required; this fusion, along with
the detection thresholds, snapping-corridor scoring, and filtering
steps, is detailed in \cref{app:inference}.

% ---- ChargeLineNet decoded outputs on example CSM images ----

% -----------------------------------------------------
\subsection{Fine-Tuning on Experimental Data}
\label{subsec:finetuning}

Synthetic training images are generated by a parameterised simulator
designed to reproduce the features of experimental CSMs as faithfully as
possible (charge-transition lines, branching from parasitic dots,
background structure, and noise).  The simulator need not be exact,
however: any residual mismatch with real measurements is what fine-tuning
is designed to close.  This residual \emph{domain gap} otherwise causes a
systematic drop in performance when a model trained only on synthetic
data is applied to real measurements.  Experimental images differ from
synthetic ones in global contrast and brightness, background structure,
scan artefacts, and device-specific features that the simulator cannot
reproduce exactly (\cref{app:domaingap} details each category).
Fine-tuning on hand-labelled experimental images closes this gap.
Only human-countable images are included in the \textsc{ChargeLineNet}
fine-tuning set, the same population used for evaluation
(\cref{subsec:lineheatmap}), since uncountable images lack reliable
ground-truth line annotations and would introduce label noise.  Training
is continued on real data with a
reduced learning rate and selectively frozen layers, so the model adapts
its feature representations to the experimental domain while retaining
the structural knowledge acquired during synthetic pre-training.  In the
experiments reported here, \textsc{ChargeLineNet} is fine-tuned on
$6{,}452$ hand-labelled CSM images (fine-tuning
hyperparameters in \cref{app:hyperparams}).  Both fine-tuning and
from-scratch training use a $95/5$ train--test split of these images to
monitor for overfitting during training; this split is internal to the
training-device data and is entirely separate from the held-out
validation devices, on which all reported accuracies are measured
(\cref{sec:model}).

We freeze the encoder: the feature-extraction backbone is
held fixed while the task-specific decoder and output head are retrained,
which balances adaptation against stability for this fine-tuning set, so
that only around a quarter of the network is adapted to the experimental
domain (\cref{app:domaingap} discusses the freeze-mode tradeoff and gives
the exact trainable-parameter count).

Note that fine-tuning applies specifically to \textsc{ChargeLineNet},
which is pre-trained on synthetic data.  The quality classifier
\textsc{CSMClassifier} (\cref{subsec:confusion_classifier}) is
trained from scratch on experimental data and does not require a
synthetic pre-training or fine-tuning stage.

On-the-fly domain-gap augmentation is additionally applied to the input
images during fine-tuning, so the model sees a wider range of artefact
combinations than the small hand-labelled dataset alone provides
(\cref{app:domaingap}).

% =====================================================
\section{Results and Discussion}
\label{sec:results}

Both models are evaluated on the 16 entirely held-out devices described
in \cref{sec:model}: \textsc{CSMClassifier} on all 2{,}407
quality-labelled images and \textsc{ChargeLineNet} on the 1{,}131-image
human-countable subset (those with hand-labelled line coordinates,
spanning 0--7 lines).  We first characterise each model on its own task
and then, in \cref{subsec:discussion}, evaluate the two chained together
as a single pipeline; this combined performance is measured on a fixed
offline dataset.

% ---------------------------------------------------------
\subsection{\textsc{CSMClassifier} Performance}%
\label{subsec:results_classifier_perf}

\begin{figure*}[t]
\centering
\includegraphics[width=0.76\textwidth]{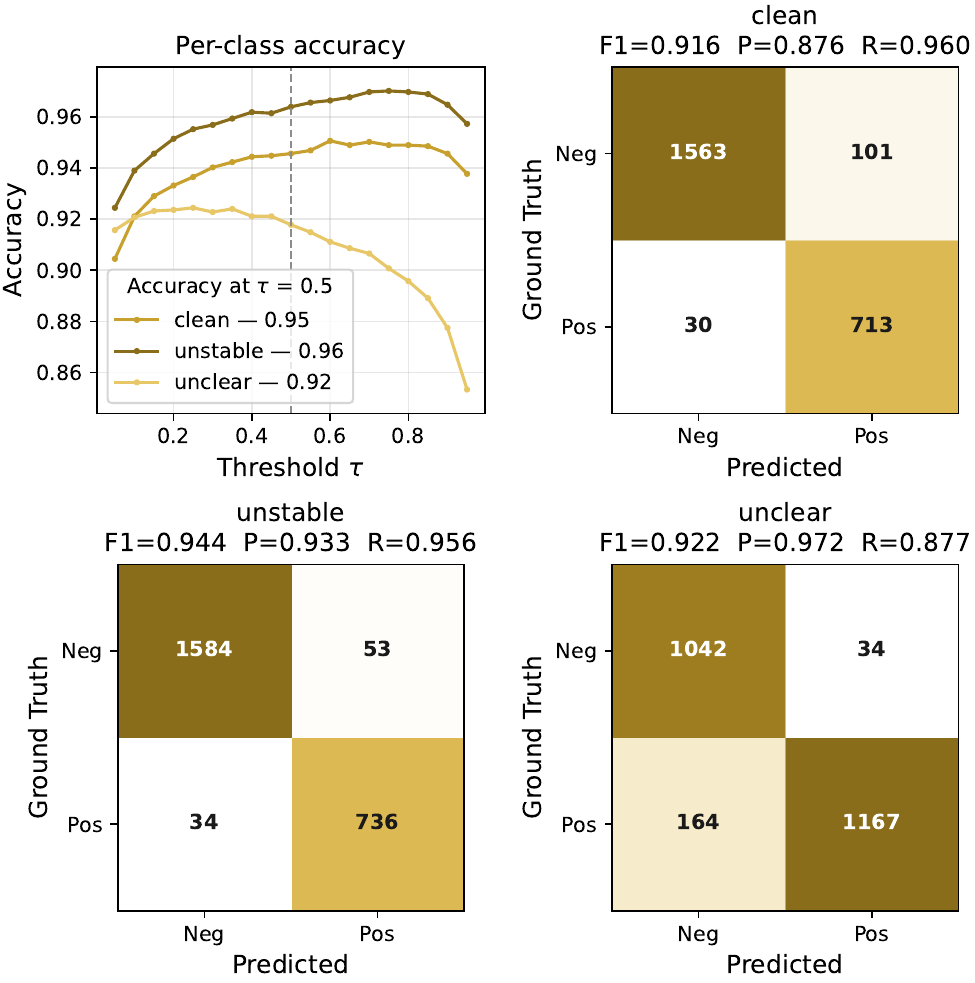}
\caption{\textsc{CSMClassifier} performance on the 2{,}407-image held-out
validation set.
(a)~Per-class accuracy as a function of the sigmoid decision threshold
$\tau$; the dashed vertical line marks $\tau = 0.5$, and the legend
reports the per-class accuracy at that threshold (clean 0.95, unstable
0.96, unclear 0.92).
(b--d)~Binary confusion matrices at $\tau = 0.5$ for the clean, unstable,
and unclear classes respectively, each annotated with its F1 score,
precision (P), and recall (R).}%
\label{fig:classification_accuracy}
\end{figure*}

On the in-device test split (5\,\% of data from the first 16
devices), \textsc{CSMClassifier} achieves a macro-averaged F1 of 0.95.
\Cref{tab:classifier_metrics_perf} reports per-class and overall
metrics on the separate 2{,}407-image validation set from the held-out 16
devices, providing a more stringent test of cross-device generalisation.
\Cref{fig:classification_accuracy} shows the corresponding
per-class accuracy as a function of the decision threshold together with
the per-class confusion matrices at $\tau = 0.5$.

\begin{table*}[t]
\centering
\caption{Multi-label classification performance of \textsc{CSMClassifier}
on 2{,}407 held-out experimental images.  Each class is evaluated
independently via a sigmoid decision threshold at $\tau = 0.5$.  Bracketed
values are 95\,\% bootstrap confidence intervals.}%
\label{tab:classifier_metrics_perf}
\begin{tabular}{lcccc}
\toprule
Class     & Precision & Recall & F1 [95\,\% CI]      & Accuracy [95\,\% CI] \\
\midrule
Clean     & 0.88      & 0.96   & 0.92 [0.90, 0.93]   & 0.95 [0.94, 0.95] \\
Unstable  & 0.93      & 0.96   & 0.94 [0.93, 0.95]   & 0.96 [0.96, 0.97] \\
Unclear   & 0.97      & 0.88   & 0.92 [0.91, 0.93]   & 0.92 [0.91, 0.93] \\
\midrule
Macro avg &           &        & 0.93 [0.92, 0.94]   & 0.94 \\
\bottomrule
\end{tabular}
\end{table*}

% ---- CSMClassifier threshold sweep and confusion matrices ----

The clean class achieves high recall (0.96) while maintaining
strong precision (0.88): only 101 of 1{,}664 non-clean images are
incorrectly labelled as clean, giving a false-positive rate of 6.1\,\%
relative to the non-clean population.  False
positives on the clean label are the most costly error because they would
feed corrupted images to the line detector.

% ---------------------------------------------------------
\subsection{\textsc{ChargeLineNet} Performance}%
\label{subsec:results_chargelinenet_perf}

\begin{figure*}[t]
\centering
\includegraphics[width=0.76\textwidth]{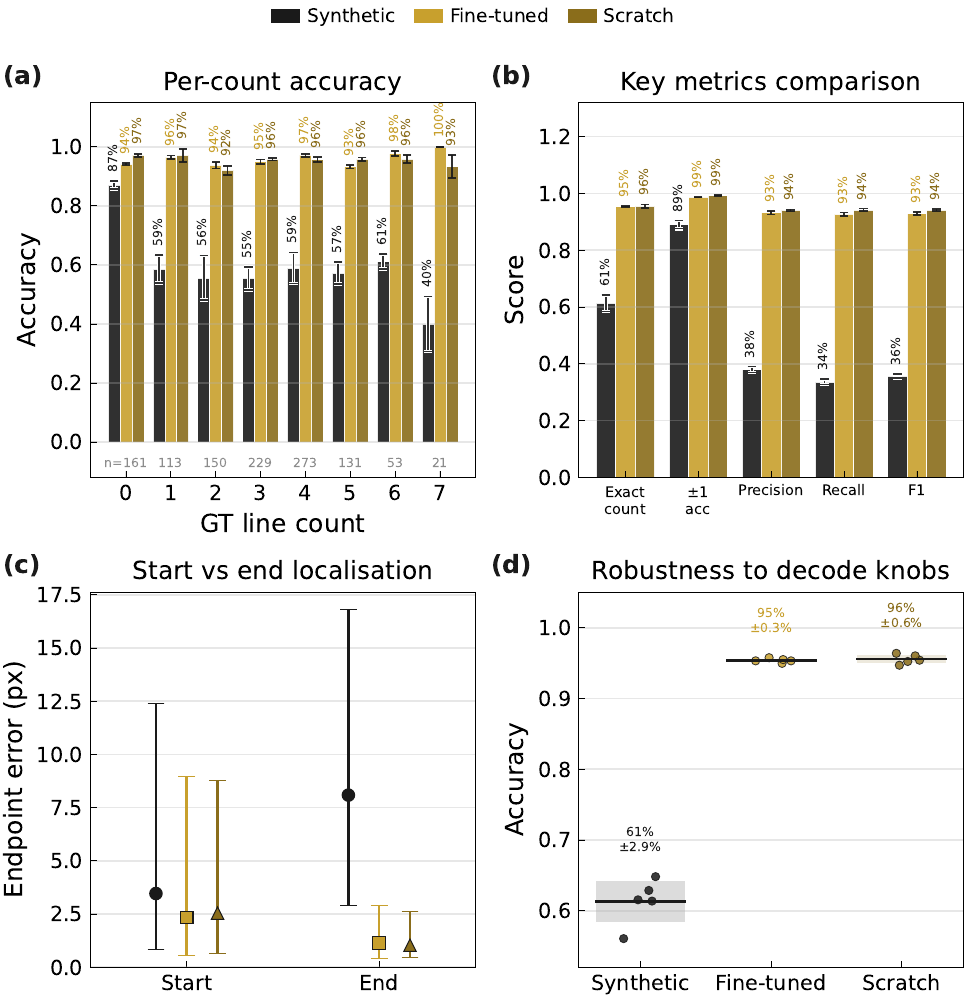}
\caption{Performance of the three \textsc{ChargeLineNet} training
variants on the 1{,}131-image held-out validation set.  The variants are:
synthetic, trained on synthetic data only; fine-tuned, i.e.\ pre-trained
on synthetic data and then adapted on experimental data with the encoder
frozen; and scratch, trained on experimental data from scratch.
(a)~Exact line-count accuracy broken down by ground-truth electron
occupancy (0--7 lines); the annotations below each group give the number
of test images in that occupancy bin.  Accuracy stays high across the full
occupancy range; the high-occupancy bins (6--7 lines) are the most sparsely
sampled, so their per-bin accuracy is estimated from few images.
(b)~Aggregate metrics (exact-count accuracy, $\pm 1$ accuracy,
line-detection precision, recall, and $F_1$).
(c)~Start- and end-point localisation error in pixels, shown as the
median with 16th--84th-percentile whiskers.  End points
are localised markedly more precisely than start points.
(d)~Robustness to the heatmap-decoding configuration: exact line-count
accuracy across five decode configurations for each variant, with the
mean and standard deviation annotated.  The trained models are insensitive
to the decode knobs (spread ${\lesssim}1\,\%$), so their accuracy is not an
artefact of a particular decoding choice.
Across all metrics the synthetic model trails substantially, while the
fine-tuned and scratch models perform comparably.}%
\label{fig:per_occupancy_accuracy}
\end{figure*}

We compare three training variants of \textsc{ChargeLineNet}:
(i)~trained on synthetic data only (``synthetic''),
(ii)~pre-trained on synthetic data and then fine-tuned on experimental
data with the encoder frozen (``fine-tuned''), and (iii)~trained from
scratch on experimental data with all layers trainable (``scratch'').
\Cref{fig:per_occupancy_accuracy} summarises their performance across
occupancy, aggregate detection metrics, endpoint localisation, and
robustness to the heatmap-decoding configuration.  The synthetic
model trails substantially on every metric, while the
fine-tuned and scratch models perform comparably.

\paragraph{Line-count accuracy.}

The primary metric is exact line-count accuracy: the fraction of test
images for which the predicted number of charge-transition lines matches
the ground-truth count.  \Cref{tab:linecount} reports this metric
for the three variants.

\begin{table}[h]
\centering
\caption{Line-count accuracy on the 1{,}131-image held-out validation set
(the human-countable subset of the held-out images, those assigned
hand-labelled line coordinates, spanning 0--7 lines) for three
\textsc{ChargeLineNet} training configurations.  ``$\pm 1$''
counts a prediction as correct if it is within one line of the
ground truth.}%
\label{tab:linecount}
\begin{tabular}{lcc}
\toprule
Configuration        & Exact (\%) & $\pm 1$ (\%) \\
\midrule
Synthetic            & 61.4       & 89.5 \\
Fine-tuned           & 95.3       & 98.6 \\
Scratch              & 96.4       & 99.4 \\
\bottomrule
\end{tabular}
\end{table}

Fine-tuning on experimental data yields 95.3\,\% exact line-count
accuracy, a roughly 34 percentage-point improvement over the synthetic
baseline; this figure is stable across the five heatmap-decoding
configurations, varying by ${\lesssim}1\,\%$
(\cref{fig:per_occupancy_accuracy}d), so it is not an artefact of a
particular decoding choice.  The $\pm 1$ accuracy of 98.6\,\% indicates that the remaining
errors are off-by-one miscounts, typically caused by faint lines at the
edge of the gate-voltage window where the charge transition is only
partially visible.  Accuracy remains high across the full range of
electron occupancies rather than degrading as the line count grows
(\cref{fig:per_occupancy_accuracy}a); the model resolves the
densely-packed transitions of the high-occupancy configurations about as
reliably as the well-separated low-occupancy ones.  The one caveat at high
occupancy is one of sampling rather than a loss of accuracy: the 6--7-line
bins are the most sparsely sampled, so the per-bin accuracy there is
estimated from few images and should be read as indicative rather than
precise.  This under-representation reflects the cost of acquiring
high-occupancy experimental CSMs rather than a weakness of the model, and
we expect the pipeline to extend gracefully to higher occupancies; we
elaborate in \cref{app:occupancy}.

\paragraph{Endpoint localisation error.}

For matched lines, end points are localised markedly more precisely than
start points (\cref{fig:per_occupancy_accuracy}c), consistent with the
design choice of anchoring the offset vector at the fading tip of each
transition line, which is visually unique and set apart from the
branching structure in which the start point is embedded; we report the
error distributions in \cref{app:inference}.

\paragraph{Selective detection of plunger-gate transitions.}

\textsc{ChargeLineNet} is trained to detect only the near-vertical
plunger-gate transitions and to ignore the branching features from
parasitic device-gate dots and other background transitions
(\cref{subsec:lineheatmap}).  The high line-detection precision on
the held-out set ($0.93$, \cref{fig:per_occupancy_accuracy}b) shows
that false detections are scarce, consistent with the model suppressing
such spurious transitions.

% ---- ChargeLineNet variant comparison ----

% ---------------------------------------------------------
\subsection{Discussion}
\label{subsec:discussion}

Having characterised \textsc{CSMClassifier} and \textsc{ChargeLineNet} on
their respective tasks, we now discuss the properties that bear on their
practical use, including their performance when chained into a single
pipeline.

% ---- Data-efficiency comparison: fine-tuned vs scratch ----
\begin{figure}[t]
\centering
\includegraphics[width=\columnwidth]{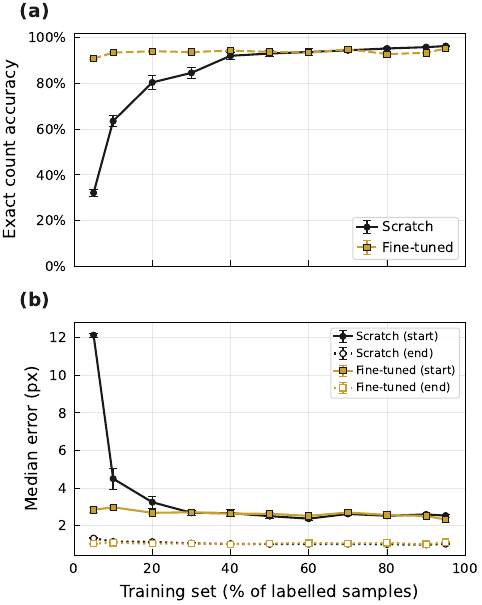}
\caption{Data efficiency of synthetic pre-training for
\textsc{ChargeLineNet}, as a function of the fraction of hand-labelled
experimental data used for training, for a model pre-trained on synthetic
data and then fine-tuned versus a model trained from scratch on
experimental data only.  Both quantities are evaluated on the held-out
experimental validation set.
(a)~Exact line-count accuracy.  With abundant labelled data the two are
comparable, but the from-scratch model degrades sharply below
${\sim}30\,\%$ of the data and collapses in the low-data regime, whereas
the fine-tuned model remains above $90\,\%$ even with only $5\,\%$ of the
labels.
(b)~Median endpoint localisation error, for line start and end points
(in pixels on $128 \times 128$ images).  The from-scratch model's
localisation degrades steeply as labels become scarce, while the
fine-tuned model's stays low throughout.  The start-point error grows
most steeply of all: the start point is embedded in the branching
structure of the charge transitions, so localising it precisely requires
a learned model of how that branching appears.  Synthetic pre-training
provides this, whereas the from-scratch model has not acquired it in the
low-data regime.
Synthetic pre-training thus provides a large gain in label efficiency in
both counting and localisation.}%
\label{fig:finetuning_comparison}
\end{figure}

\paragraph{Domain gap and fine-tuning.}

The roughly 34 percentage-point improvement from synthetic ($61.4$\%) to fine-tuned ($95.3$\%) demonstrates that the domain gap between simulated and experimental CSM images is substantial but can be closed with a moderate amount of hand-labelled data. The magnitude of this gain is task- and pipeline-dependent: \citet{darulova2021evaluation}, studying charge-state classification, report that transfer learning from a synthetic pre-trained network improves accuracy on ideal experimental data by up to only ${\sim}5\,\%$, and that merging synthetic and experimental data in a single training step was more effective for them than freezing layers. That benefit, however, was measured with their full experimental training set available; they did not vary the amount of experimental data, and identified adaptation under scarce data as a promising but untested direction. Consistent with their full-data result, we too find that when labels are plentiful our fine-tuned and from-scratch models are essentially tied. The value of synthetic pre-training instead emerges most clearly through its \emph{label efficiency}, revealed only when the amount of hand-labelled data is varied down into the low-data regime.
\Cref{fig:finetuning_comparison} compares the fine-tuned model
against a model trained from scratch on experimental data only, as the
amount of hand-labelled experimental data is varied.  When the full
labelled set is available the two are comparable ($95.3\,\%$ fine-tuned
versus $96.4\,\%$ from scratch), but their behaviour diverges sharply as
labels become scarce: the from-scratch model degrades rapidly below
${\sim}30\,\%$ of the data and collapses in the low-data
regime ($32.1\,\%$ exact-count accuracy at $5\,\%$ of the labels),
whereas the fine-tuned model remains above $90\,\%$ throughout, still
reaching $90.9\,\%$ at that same $5\,\%$ of the labels
(\cref{fig:finetuning_comparison}a).  The same pattern holds for endpoint
localisation (\cref{fig:finetuning_comparison}b): the from-scratch model's
localisation error grows steeply as labels become scarce, while the
fine-tuned model's stays low throughout.  This
dictates when each training route is preferable in practice: where a large
labelled set already exists, as is the case here, training from scratch is
sufficient and synthetic pre-training is not required to reach peak
accuracy, but its value shows when labels are scarce, as when adapting
the pipeline to a new device generation, sensor design, or artefact
regime.  There, synthetic pre-training followed by
fine-tuning reaches usable accuracy with a fraction of the labelling
effort, a decisive advantage given that hand-labelling charge-transition
lines is the principal bottleneck in adapting the models to unseen device
families, and is the route we recommend for bootstrapping them.

The point-based labelling strategy compounds this advantage: because
\textsc{ChargeLineNet} is supervised only on the start and end
coordinates of each line (\cref{subsec:lineheatmap}), annotating
an image requires just two clicks per transition rather than a dense
per-pixel mask, so each of the already-fewer labels is itself far cheaper
to produce.  This lightweight annotation format is also well suited to
future automation by a vision-language model, which we discuss under
\nameref{par:limitations} below as a route to both reducing labelling
effort and improving annotator consistency.

The fine-tuned model's accuracy is nearly flat between
${\sim}20\,\%$ and $100\,\%$ of the experimental data, indicating that
beyond a modest amount of labelled data further gains come not from the
sheer quantity of examples but from the \emph{variety} of devices,
cooldowns, and artefact types represented.  We expect accuracy
to improve most when the labelled set is broadened to cover new device
geometries, sensor behaviours, and failure modes rather than simply
enlarged.

A further advantage is that the synthetic images themselves are cheap to
produce.  Rather than solving the electrostatics and quantum transport of
each device from first principles, our generator synthesises the salient
image features (transition lines and their branching structure, spurious
transitions, background structure, and noise) directly
(\cref{app:synthetic}), so the ${>}10^5$ pre-training images are
produced at negligible cost and the corpus can be regenerated cheaply
whenever the target distribution changes.

\paragraph{End-to-end pipeline.}

In deployment the two models are chained so that \textsc{CSMClassifier}
gates \textsc{ChargeLineNet} (\cref{sec:model}), using the fine-tuned
\textsc{ChargeLineNet} variant as the line counter.  Evaluated on the held-out
images common to both validation sets, the clean-image quality gate (the
\textsc{CSMClassifier} clean output at $\tau = 0.5$) passes
$96\,\%$ of truly clean images, matching its clean-class recall in
\cref{tab:classifier_metrics_perf}, and the line counter then reads their
occupancy exactly on $97.9 \pm 0.2\,\%$ of those it receives, so the
combined pipeline returns the correct electron count on
$93.8 \pm 0.2\,\%$ of clean held-out images end-to-end (mean and standard
deviation across five heatmap-decoding configurations).  The end-to-end
figure is thus the product of the two stages ($0.96 \times 0.979$): it
sits below the counter's own $97.9\,\%$ because the ${\sim}4\,\%$ of clean
images the gate wrongly rejects are scored as pipeline misses even though
the counter never sees them.  Counting the
degraded images that the gate correctly rejects for re-measurement, the
full pipeline handles $93.9 \pm 0.05\,\%$ of all held-out images correctly.
The line counter's accuracy on the gate-passed images
($97.9\,\%$) is slightly higher than its standalone accuracy on the full
validation set ($95.3\,\%$, \cref{tab:linecount}).  This is expected: the
standalone set is defined by human \emph{countability} rather than
cleanliness, and so includes mildly-degraded images that are harder for
the line detector, whereas the gate filters these out and forwards only
images labelled clean, on which the counter is more reliable.
That the end-to-end accuracy matches the product of the two stages'
independent accuracies indicates that the two models' errors are largely
uncorrelated and that the gate seldom discards images the counter would
have read correctly.

\paragraph{Inference speed and deployment.}

Both models are lightweight and fast.  A forward pass requires only
$3.6$\,GFLOP per $128 \times 128$ image.  On a GPU (NVIDIA GH200), a
forward pass completes in ${\sim}0.3$\,ms per image.  Even without a GPU, on
a consumer-grade CPU (Intel Core Ultra 7 165U, 2.10\,GHz) a forward pass
through \textsc{CSMClassifier} takes under 10\,ms and
\textsc{ChargeLineNet} takes under 50\,ms (forward pass only, the decoding algorithm consumes a further ${\sim}2$\,ms). The
on-disk footprint is 2.5\,MB for \textsc{CSMClassifier} and 4\,MB for
\textsc{ChargeLineNet}, a combined 6.5\,MB, so neither model is the
bottleneck in a measure--infer--adjust tuning cycle and both deploy on
standard laboratory hardware with or without a GPU.

\paragraph{Limitations.}
\label{par:limitations}

Several limitations should be noted.  First, the validation set is
drawn from a limited number of devices.  Second, the end-to-end pipeline
accuracy reported above is measured purely on a fixed offline dataset: each
pre-recorded image is passed through the gate and counter once, and an
image the quality gate flags is counted as a pipeline error even though
this is a recoverable action that costs measurement time rather than
readout accuracy.  In practice the pipeline would sit inside a live
measure--infer--decide loop in which flagged images prompt the operator to
retune and re-measure rather than being discarded, so the accuracy realised
in deployment would differ from the offline numbers reported here.  Finally, the
hand-labelled training data
introduces annotator subjectivity, particularly for borderline images
where the distinction between clean, unstable, and unclear is ambiguous.
Even with written labelling rules for reference, hand-labelling remains
prone to mistakes and to inconsistency that accumulates over the long
labelling sessions required to annotate thousands of images; inter-annotator
agreement has not been formally quantified.  A promising route to mitigating
this, which we do not pursue in the present work, is vision-language-model
(VLM) assisted annotation.  The lightweight, point-based label format of
\textsc{ChargeLineNet} is well suited to it: a VLM could propose start/end
coordinates directly, reducing the human role to verification rather than
exhaustive labelling, and the same approach could assign
\textsc{CSMClassifier}'s per-image quality labels (clean, unstable, unclear)
in place of a human annotator.  Recent work benchmarking vision-language
models on quantum calibration plots~\cite{Cao2026-qcaleval} suggests such
assisted annotation is increasingly feasible.  Beyond reducing labelling
effort, a VLM applies the same labelling criteria uniformly across an
arbitrarily large set without fatigue or drift, and can serve as a consistent
second annotator against which human labels are cross-checked and
inter-annotator agreement quantified, directly addressing the consistency
concern above.

% =====================================================
\section{Conclusion}
\label{sec:conclusion}

We have presented two compact CNN models for automated analysis of
charge stability maps from double-quantum-dot devices operated in the
isolated-mode regime.  \textsc{CSMClassifier} screens each measured CSM
for quality, identifying the well-resolved images suitable for line
detection, while \textsc{ChargeLineNet} localises charge-transition lines
via signed-heatmap and offset-vector regression, counting only the
plunger-gate transitions that set the electron occupancy while ignoring
the branching structure from parasitic dots and spurious background
transitions, to read off the electron occupancy of the images that pass.
Both models are validated on 16 devices held out entirely from training
and scored against hand-labelled ground-truth labels: on this unseen
data \textsc{CSMClassifier} attains 94\,\% macro-averaged accuracy across
three quality categories and \textsc{ChargeLineNet} reaches 95.3\,\%
exact line-count accuracy, demonstrating cross-device generalization.
Combined into a single pipeline, in which the classifier first screens
each image and passes only those judged clean to the fine-tuned line
counter, the two models together return the correct electron occupancy for
$93.8\,\%$ of clean held-out images, and handle $93.9\,\%$ of
all held-out images correctly once the correct rejection of degraded ones
is included.

For \textsc{ChargeLineNet}, this accuracy can be reached either by
training from scratch on a few thousand hand-labelled images or by
fine-tuning a model pre-trained on cheap, non-physics-based synthetic
data, showing that the domain gap between simulated and real CSMs can be
closed either way.  Fine-tuning is nonetheless the more practical route:
it retains above 90\,\% accuracy when hand-labelled data is scarce, where
training from scratch collapses, and is therefore the route we recommend
for bootstrapping the models on a new device generation.

Both models are compact (6.5\,MB combined) and fast enough to run on
standard laboratory hardware with or without a GPU.  Together they bring
automated CSM analysis, so far developed for the reservoir-coupled
regime, to the increasingly adopted isolated mode of operation,
addressing a key step towards the scalable, reproducible device tuneup
that fault-tolerant spin-qubit processors will demand.

% =====================================================
\begin{acknowledgments}
We thank the imec Quantum Computing team and the qprobe team for their 
support. We acknowledge support from the Australian Research Council 
(FL190100167, IE240100252, IM230100396) the U.S. Army Research Office
(W911NF-23-10092), and the U.S. Air Force Office of Scientific Research
(FA2386-22-1-4070). M.C. acknowledges support from the Sydney Quantum Academy.
\end{acknowledgments}
         
% =====================================================
% =====================================================
\appendix

% Make cleveref call appendix sections/subsections "Appendix" rather than
% "section", so they are not confused with main-body sections.  \crefalias
% retypes only labels defined after \appendix, leaving main-body \cref intact.
\crefalias{section}{appendix}
\crefalias{subsection}{appendix}
\crefname{appendix}{Appendix}{Appendices}
\Crefname{appendix}{Appendix}{Appendices}

\section{Dataset Splits}
\label{app:splits}

The image counts quoted throughout the text all derive from the same 32
devices, partitioned into 16 training devices and 16 held-out devices, with
the two models drawing on different image populations of that partition
according to their task.  \Cref{tab:splits} consolidates these counts in
one place.  \textsc{CSMClassifier} is trained and evaluated on all
quality-labelled images.  \textsc{ChargeLineNet} uses the narrower
human-countable subset (those images to which a reliable line count, and
hence hand-labelled start/end coordinates, could be assigned; spanning the
0--7-line occupancy range), since only these carry the ground-truth line
annotations it requires.  Both held-out sets come from the same 16 held-out
devices, so the 1{,}131-image line-count set is the human-countable subset of
the 2{,}407-image quality-screening set.

\begin{table}[h]
\centering
\caption{Dataset splits.  All counts derive from the same 16 training / 16
held-out device partition.  \textsc{CSMClassifier} uses all quality-labelled
images; \textsc{ChargeLineNet} uses the human-countable subset that carries
hand-labelled line coordinates.  The training-device images are further
divided by a random 95/5 train/test split within those 16 devices.}%
\label{tab:splits}
\begin{tabular}{llcc}
\toprule
Model & Split & Images & Devices \\
\midrule
\textsc{CSMClassifier} & Training (labelled)   & 10{,}060 & 16 \\
                       & Held-out (validation) &  2{,}407 & 16 \\
\midrule
\textsc{ChargeLineNet} & Fine-tuning (labelled) & 6{,}452 & 16 \\
                       & Held-out (validation)  & 1{,}131 & 16 \\
\bottomrule
\end{tabular}
\end{table}

\section{Architecture Details}
\label{app:architecture}

\begin{figure*}[t]
\centering
\resizebox{\textwidth}{!}{%
\begin{tikzpicture}[
  node distance=0.16cm and 0.10cm,
  block/.style={rectangle, rounded corners=2.5pt, draw=black!65, line width=0.4pt,
                fill=#1, text width=1.28cm, align=center,
                minimum height=1.05cm, font=\scriptsize},
  block/.default=blue!12,
  arrow/.style={-{Stealth[length=3.2pt]}, thick, draw=black!70},
  dimlab/.style={font=\scriptsize, text=black!45, inner sep=1pt},
  grouplab/.style={font=\scriptsize\scshape, text=black!55},
  groupbox/.style={rounded corners=4pt, inner xsep=4pt, inner ysep=6pt},
]

  % ── Blocks ─────────────────────────────────────────
  \node[block=gray!15]   (input)
        {\textbf{Input}\\$1{\times}128^2$};
  \node[block=orange!22, right=of input]  (stem)
        {\textbf{Stem}\\LCN $\|$ IN\\$3{\times}3$ conv\\$\to 16$ ch};
  \node[block=blue!15,   right=of stem]   (enc1)
        {\textbf{Enc\,1}\\$3{\times}3$ res\\$16{\to}16$\\Pool$\,\downarrow\!2$};
  \node[block=blue!20,   right=of enc1]   (enc2)
        {\textbf{Enc\,2}\\$3{\times}3$ res\\$16{\to}32$\\Pool$\,\downarrow\!2$};
  \node[block=blue!25,   right=of enc2]   (enc3)
        {\textbf{Enc\,3}\\$3{\times}3$ res\\$32{\to}64$\\Pool$\,\downarrow\!2$};
  \node[block=blue!30,   right=of enc3]   (enc4)
        {\textbf{Enc\,4}\\$3{\times}3$ res\\$64{\to}128$\\Pool$\,\downarrow\!2$};
  \node[block=purple!20, right=of enc4]   (bn)
        {\textbf{Bneck}\\$3{\times}3$ res\\128\,ch\\Drop2d\,0.1};
  \node[block=red!18,    right=of bn]     (gap)
        {\textbf{GAP}\\Adaptive\\AvgPool\\$\to128{\times}1^2$};
  \node[block=red!15,    right=of gap]    (head)
        {\textbf{MLP Head}\\Drop\,0.1\\$128{\to}128$\\Drop\,0.05\\$128{\to}3$};
  \node[block=yellow!35, right=of head]   (output)
        {\textbf{Output}\\sigmoid\\clean\\unstable\\unclear};

  % ── Forward arrows ─────────────────────────────────
  \foreach \a/\b in {input/stem, stem/enc1, enc1/enc2, enc2/enc3,
                     enc3/enc4, enc4/bn, bn/gap, gap/head, head/output}
    \draw[arrow] (\a) -- (\b);

  % ── Grouping bands (background) ─────────────────────
  \begin{scope}[on background layer]
    \node[groupbox, fill=blue!7,   draw=blue!30,   fit=(enc1)(enc4)] (encgrp) {};
    \node[groupbox, fill=red!6,    draw=red!30,    fit=(gap)(head)]  (headgrp) {};
  \end{scope}

  % ── Spatial dimension annotations ──────────────────
  \node[dimlab, below=2pt of encgrp.south -| stem]  {$128^2$};
  \node[dimlab, below=2pt of encgrp.south -| enc1]  {$64^2$};
  \node[dimlab, below=2pt of encgrp.south -| enc2]  {$32^2$};
  \node[dimlab, below=2pt of encgrp.south -| enc3]  {$16^2$};
  \node[dimlab, below=2pt of encgrp.south -| enc4]  {$8^2$};
  \node[dimlab, below=2pt of encgrp.south -| bn]    {$8^2$};
  \node[dimlab, below=2pt of encgrp.south -| gap]   {$1^2$};

  % ── Group titles (below dim row) ───────────────────
  \node[grouplab, below=11pt of encgrp.south]  {Encoder};
  \node[grouplab, below=11pt of headgrp.south] {Classifier head};

\end{tikzpicture}%
}
\caption{\textsc{CSMClassifier} encoder-only architecture (see
\cref{subsec:confusion_classifier}).  Spatial dimensions after
each stage are shown below each block.  The histogram-invariant stem
(LCN $\|$ InstanceNorm) is identical to \textsc{ChargeLineNet}.  The stem projects to 16 channels.  Four
encoder stages use isotropic $3{\times}3$ residual blocks (two
convolutions with batch normalisation, ReLU, and a $1{\times}1$
projection shortcut) followed by $2{\times}2$ max-pooling, progressively
increasing the channel width from 16 to 128.  The bottleneck applies a
residual block with spatial dropout ($p{=}0.1$).  Adaptive global
average pooling (GAP) collapses the $8{\times}8$ feature map to a single
128-dimensional vector.  A two-layer MLP head with dropout ($p{=}0.1$
then $p{=}0.05$) maps this vector to three class logits (clean,
unstable, unclear); independent sigmoid activations and
thresholding determine active labels.  Binary cross-entropy with
logits loss and optional per-class positive weights handles class
imbalance.  No decoder or skip connections are needed because the task
requires image-level labels.}%
\label{fig:classifier_flow}
\end{figure*}
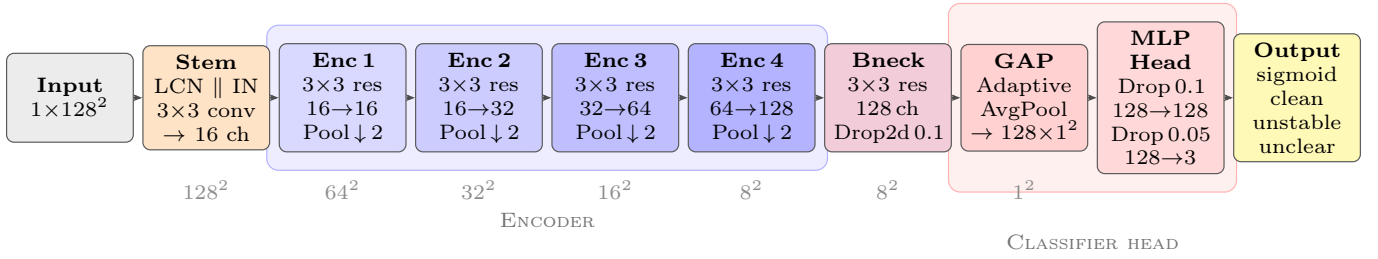

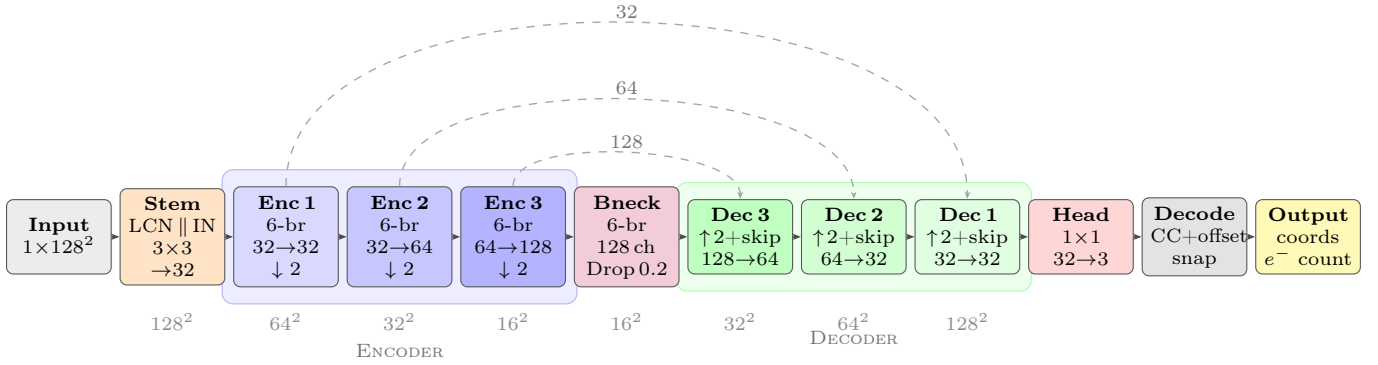
\begin{figure*}[t]
\centering
\resizebox{\textwidth}{!}{%
\begin{tikzpicture}[
  node distance=0.16cm and 0.10cm,
  block/.style={rectangle, rounded corners=2.5pt, draw=black!65, line width=0.4pt,
                fill=#1, text width=1.12cm, align=center,
                minimum height=0.95cm, font=\scriptsize},
  block/.default=blue!12,
  arrow/.style={-{Stealth[length=3.2pt]}, thick, draw=black!70},
  skip/.style={-{Stealth[length=2.8pt]}, dashed, draw=gray!70, line width=0.5pt},
  skiplab/.style={font=\scriptsize, text=black!55, inner sep=1pt},
  dimlab/.style={font=\scriptsize, text=black!45, inner sep=1pt},
  grouplab/.style={font=\scriptsize\scshape, text=black!55},
  groupbox/.style={rounded corners=4pt, inner xsep=4pt, inner ysep=6pt},
]

  % ── Blocks ─────────────────────────────────────────
  \node[block=gray!15]   (input)
        {\textbf{Input}\\$1{\times}128^2$};
  \node[block=orange!22, right=of input]  (stem)
        {\textbf{Stem}\\LCN$\,\|\,$IN\\$3{\times}3$\\${\to}32$};
  \node[block=blue!15,   right=of stem]   (enc1)
        {\textbf{Enc\,1}\\6-br\\$32{\to}32$\\$\downarrow\!2$};
  \node[block=blue!22,   right=of enc1]   (enc2)
        {\textbf{Enc\,2}\\6-br\\$32{\to}64$\\$\downarrow\!2$};
  \node[block=blue!30,   right=of enc2]   (enc3)
        {\textbf{Enc\,3}\\6-br\\$64{\to}128$\\$\downarrow\!2$};
  \node[block=purple!20, right=of enc3]   (bn)
        {\textbf{Bneck}\\6-br\\128\,ch\\Drop\,0.2};
  \node[block=green!25,  right=of bn]     (dec3)
        {\textbf{Dec\,3}\\$\uparrow\!2$+skip\\$128{\to}64$};
  \node[block=green!18,  right=of dec3]   (dec2)
        {\textbf{Dec\,2}\\$\uparrow\!2$+skip\\$64{\to}32$};
  \node[block=green!12,  right=of dec2]   (dec1)
        {\textbf{Dec\,1}\\$\uparrow\!2$+skip\\$32{\to}32$};
  \node[block=red!16,    right=of dec1]   (head)
        {\textbf{Head}\\$1{\times}1$\\$32{\to}3$};
  \node[block=gray!22,   right=of head]   (decode)
        {\textbf{Decode}\\CC+offset\\snap};
  \node[block=yellow!35, right=of decode] (output)
        {\textbf{Output}\\coords\\$e^-$ count};

  % ── Forward arrows ─────────────────────────────────
  \foreach \a/\b in {input/stem, stem/enc1, enc1/enc2, enc2/enc3,
                     enc3/bn, bn/dec3, dec3/dec2, dec2/dec1,
                     dec1/head, head/decode, decode/output}
    \draw[arrow] (\a) -- (\b);

  % ── Skip connections (arced above) ─────────────────
  \draw[skip] (enc1.north) to[out=90,in=90,looseness=0.85]
        node[skiplab, above=0.5pt, pos=0.5] {32} (dec1.north);
  \draw[skip] (enc2.north) to[out=90,in=90,looseness=0.70]
        node[skiplab, above=0.5pt, pos=0.5] {64} (dec2.north);
  \draw[skip] (enc3.north) to[out=90,in=90,looseness=0.60]
        node[skiplab, above=0.5pt, pos=0.5] {128} (dec3.north);

  % ── Grouping bands (background) ─────────────────────
  \begin{scope}[on background layer]
    \node[groupbox, fill=blue!7,  draw=blue!30,  fit=(enc1)(enc3)] (encgrp) {};
    \node[groupbox, fill=green!7, draw=green!35, fit=(dec3)(dec1)] (decgrp) {};
  \end{scope}

  % ── Spatial dimension annotations ──────────────────
  \node[dimlab, below=2pt of encgrp.south -| stem]  {$128^2$};
  \node[dimlab, below=2pt of encgrp.south -| enc1]  {$64^2$};
  \node[dimlab, below=2pt of encgrp.south -| enc2]  {$32^2$};
  \node[dimlab, below=2pt of encgrp.south -| enc3]  {$16^2$};
  \node[dimlab, below=2pt of encgrp.south -| bn]    {$16^2$};
  \node[dimlab, below=2pt of encgrp.south -| dec3]  {$32^2$};
  \node[dimlab, below=2pt of encgrp.south -| dec2]  {$64^2$};
  \node[dimlab, below=2pt of encgrp.south -| dec1]  {$128^2$};

  % ── Group titles ────────────────────────────────────
  \node[dimlab] (dimrow) at (encgrp.south -| enc2) {};
  \node[grouplab, below=11pt of encgrp.south] {Encoder};
  \node[grouplab, below=11pt of decgrp.south] {Decoder};

\end{tikzpicture}%
}
\caption{\textsc{ChargeLineNet} encoder--decoder architecture (see
\cref{subsec:lineheatmap}).  Spatial dimensions after each stage
are shown below each block.  The histogram-invariant stem (LCN $\|$
InstanceNorm) removes global contrast variation.  Three encoder levels
use six-branch anisotropic blocks ($7{\times}1$, $5{\times}5$,
$1{\times}7$, $3{\times}3$, dilated $3{\times}3$, $1{\times}15$) with
DropPath ($p{=}0.1$) and $2{\times}2$ max-pooling.  The bottleneck adds
Dropout2d ($p{=}0.2$).  Three decoder levels upsample bilinearly,
concatenate encoder skip connections, and apply two $3{\times}3$
convolutions with spatial dropout ($p{=}0.1$).  The $1{\times}1$ head
produces three unbounded channels: signed heatmap (ch\,0, $+$start /
$-$end) and offset vectors (ch\,1--2, $\Delta x$/$\Delta y$ from end
$\to$ start).  Post-processing detects end-point blobs via
connected-component labelling, averages the offset over each blob, snaps
to the nearest start blob, and filters by length and angle.  Total:
$935{,}283$ parameters.}%
\label{fig:chargelinenet_flow}
\end{figure*}

This appendix gives the layer-level specifications of the two models
whose conceptual structure is described in \cref{sec:model} and
illustrated on real data in \cref{fig:classifier_arch,fig:chargelinenet_arch}.  The corresponding schematic block
diagrams are given in \cref{fig:classifier_flow,fig:chargelinenet_flow}.

% ---- CSMClassifier flowchart ----

% ---- ChargeLineNet flowchart ----

\subsection{Shared Histogram-Invariant Stem}
\label{app:stem}

Both models begin with the same histogram-invariant stem, which renders
the network insensitive to global brightness and contrast differences
between training (simulated) and test (experimental) data.  Two parallel
branches process the single input channel:
\begin{enumerate}
    \item \textbf{Local contrast normalisation (LCN).}  A fixed
          $9\times9$ Gaussian kernel ($\sigma=1.5$\,px) blurs the input;
          the blurred image is subtracted from the original and the
          result is divided by the local standard deviation.  This
          removes any DC offset and gain variation while preserving
          edge structure.
    \item \textbf{Instance normalisation.}  \texttt{InstanceNorm2d}
          normalises each image independently to zero mean and unit
          variance, providing a complementary global contrast
          normalisation.
\end{enumerate}
The two branch outputs are concatenated and projected via a $3\times3$
convolution followed by batch normalisation and ReLU.  The only
difference between the two models is the projection width: $c_1 = 16$
channels for \textsc{CSMClassifier} and $c_1 = 32$ for
\textsc{ChargeLineNet}.

\subsection{\textsc{CSMClassifier}}
\label{app:classifier_arch}

The encoder consists of four stages.  Each stage contains a residual
convolutional block~\cite{he2016resnet} (two $3\times3$ convolutions with batch
normalisation and ReLU, plus a $1\times1$ projection shortcut when the
channel width changes) followed by $2\times2$ max-pooling, with channel
progression $[16, 32, 64, 128]$.  After the four encoder stages, a
bottleneck block (residual convolution with spatial dropout $p = 0.1$)
is followed by adaptive average pooling that collapses the spatial
dimensions to a single 128-dimensional feature vector.  A two-layer MLP
head maps this vector to three output logits: Linear($128 \to 128$) with
ReLU and dropout ($p = 0.05$), followed by Linear($128 \to 3$); dropout
before the first linear layer is set to $p = 0.1$.

\subsection{\textsc{ChargeLineNet}}
\label{app:branches}

\begin{figure*}[t]
\centering
\includegraphics[width=\textwidth]{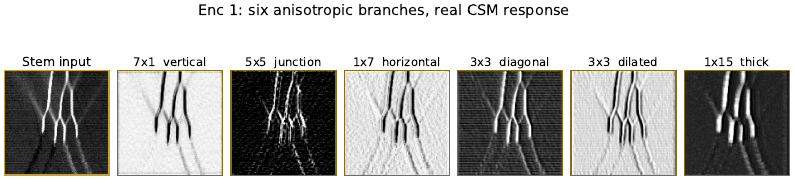}
\caption{Response of the six anisotropic branches in the first
\textsc{ChargeLineNet} encoder block to the same real CSM image (only the
single highest-variance, i.e.\ most active, channel of each branch is
shown; the remaining channels are omitted to avoid clutter).  Each branch applies
a differently shaped kernel: $7{\times}1$ (vertical), $5{\times}5$
(junction), $1{\times}7$ (horizontal), $3{\times}3$ (diagonal),
dilated $3{\times}3$, and $1{\times}15$ (thick), so that the block
responds selectively to line features across orientations and scales
(\cref{app:branches}).  The near-vertical charge-transition
lines are picked out most strongly by the tall $7{\times}1$ and
diagonal $3{\times}3$ branches.}%
\label{fig:chargelinenet_branches}
\end{figure*}

\paragraph{Anisotropic encoder.}
Each anisotropic block in the encoder processes its input through six
parallel convolutional branches whose outputs are concatenated, a
multi-branch design in the spirit of the Inception
module~\cite{szegedy2015googlenet} but with \emph{anisotropic} kernels
selective for the near-vertical charge-transition lines, rather than
Inception's multi-scale square kernels.  The six branches are designed to
detect lines at multiple orientations and scales:
\begin{itemize}
    \item $(7\times1)$: long vertical extent, targeting the dominant
          near-vertical transition lines.
    \item $(5\times5)$: junction and branching-point detection.
    \item $(1\times7)$: horizontal context, capturing gate-voltage
          modulation along the horizontal axis.
    \item $(3\times3)$: diagonal and general structure.
    \item $(3\times3)$ dilated ($d=2$): wider receptive field without
          additional parameters.
    \item $(1\times15)$: thick-line coverage for broad transitions.
\end{itemize}
All six branch outputs are concatenated and projected to the target
channel width via a $1\times1$ convolution.  A residual shortcut~\cite{he2016resnet}
(identity or $1\times1$ projection) is added before the final ReLU to
stabilise training.  Stochastic depth (DropPath, $p=0.1$) is applied
during training to prevent over-reliance on any single orientation
branch.  \Cref{fig:chargelinenet_branches} shows the response of
each of the six branches to a real CSM, confirming that they specialise
to different orientations and scales as intended.

% ---- ChargeLineNet anisotropic branch decomposition ----

\paragraph{Bottleneck, decoder, and head.}
A fourth anisotropic block operates at the coarsest resolution
($c_3=128$ channels) and is followed by \texttt{Dropout2d} ($p=0.2$) to
regularise the most abstract feature representations.  The decoder
mirrors the encoder through three upsampling blocks, each of which
bilinearly upsamples the incoming feature map, concatenates the
corresponding encoder skip connection (U-Net style), and applies two
$3\times3$ convolutions with batch normalisation, ReLU, and spatial
dropout ($p=0.1$).  Two convolutions per block are used because accurate
coordinate prediction requires spatially precise feature maps that a
single convolution cannot reconstruct after upsampling.  A single
$1\times1$ convolution then maps the full-resolution decoder features to
the three output channels with no activation function; its bias is
initialised to zero so that early predictions are near-zero everywhere.

% -----------------------------------------------------
\section{Shared Design and Architectural Differences}
\label{subsec:model_comparison}

The two models share several design choices but differ where the tasks
demand it.  \Cref{tab:pipeline_comparison} summarises the key
similarities and distinctions.

\begin{table*}[ht]
\centering
\caption{Shared and distinct design elements of the two models.}%
\label{tab:pipeline_comparison}
\begin{tabular}{lll}
\toprule
Shared                          & \textsc{CSMClassifier}              & \textsc{ChargeLineNet} \\
\midrule
Histogram-invariant stem        & Isotropic $3{\times}3$ residual     & 6-branch anisotropic blocks \\
\quad (LCN + InstanceNorm)      & \quad blocks                        & \\
                                & $c_1{=}16$; stages $[16,32,64,128]$ & $c_1{=}32$; stages $[32,64,128]$ \\
AdamW optimiser                 & AdaptiveAvgPool $\to$ MLP head      & U-Net decoder + $1{\times}1$ head \\
Cosine LR + linear warm-up     & Output: 3-class logits (sigmoid)    & Output: heatmap + offset $(3{\times}H{\times}W)$ \\
Gradient clipping (max norm 1)  & BCE with logits loss                & Weighted MSE + masked smooth-$\ell_1$ \\
Same augmentation suite         & $619{,}715$ parameters               & $935{,}283$ parameters \\
\bottomrule
\end{tabular}
\end{table*}

The shared histogram-invariant stem ensures that both models receive
feature representations that are independent of global brightness and
contrast, so a single preprocessing routine serves both models.
The base channel width differs between the two models ($c_1 = 16$ for
\textsc{CSMClassifier}, $c_1 = 32$ for \textsc{ChargeLineNet}),
reflecting the lower capacity needed for image-level classification
versus dense spatial prediction.  The
encoder blocks differ because \textsc{ChargeLineNet} requires
orientation-selective receptive fields to detect near-vertical
charge-transition lines, whereas \textsc{CSMClassifier} recognises
spatially diffuse, orientation-agnostic artefacts for which standard
isotropic convolutions suffice.  The output heads reflect the task
difference: \textsc{CSMClassifier} collapses spatial information via
global average pooling to produce an image-level label, while
\textsc{ChargeLineNet} preserves full spatial resolution through a
U-Net decoder with skip connections to localise line endpoints to
within a few pixels.

Both parameter counts are small by modern CNN standards (routinely
$>10^7$).  Low parameter counts reduce overfitting risk when
hand-labelled data is scarce and keep inference fast on standard
hardware.  Inference latency,
memory footprint, and deployment considerations are quantified in
\cref{subsec:discussion}.

\paragraph{Preprocessing.}

Both models receive images resized to $128 \times 128$ pixels and
min-max normalised to $[0, 1]$.  \textsc{CSMClassifier} additionally
subtracts the per-image mean before normalisation.
\textsc{ChargeLineNet} omits this step because its histogram-invariant
stem already removes global offsets analytically via the LCN and
instance-normalisation branches.  Apart from this single flag, the
preprocessing routine is identical for both models, and the training
data for both models are min-max normalised.

\section{Training Loss Functions}
\label{app:loss}

\subsection{\textsc{CSMClassifier}}
\label{app:loss_classifier}

The quality classifier is trained with binary cross-entropy with logits
loss applied independently to each of the three output classes:
\begin{equation}
    \mathcal{L}_{\mathrm{BCE}} =
    -\frac{1}{3}\sum_{k=0}^{2} \Bigl[
        w_k \, y_k \log\sigma(\hat{z}_k) +
        (1 - y_k)\log\bigl(1 - \sigma(\hat{z}_k)\bigr)
    \Bigr],
\end{equation}
where $\hat{z}_k$ are the raw logits, $y_k \in \{0, 1\}$ is the
multi-hot target for class $k$, $\sigma$ denotes the sigmoid function,
and $w_k$ are optional per-class positive weights computed from the
inverse positive-class frequency to handle class imbalance.  Following the
standard positive-weighting convention, $w_k$ multiplies only the
positive ($y_k = 1$) term, up-weighting the loss contribution of the
rarer positive examples while leaving the negative term unweighted; the
asymmetry is therefore by design rather than an omission.  Label
smoothing, which softens the targets $1 \to 1 - \varepsilon/2$,
$0 \to \varepsilon/2$, is supported but was disabled ($\varepsilon = 0$)
for the results reported here.

\subsection{\textsc{ChargeLineNet}}
\label{app:loss_chargelinenet}

The line detector is trained with a combined endpoint--offset loss that
supervises all three output channels jointly:
\begin{equation}
    \mathcal{L} = \mathcal{L}_{\mathrm{heat}} +
                  \lambda_{\mathrm{off}}\,\mathcal{L}_{\mathrm{off}} .
\end{equation}

The heatmap term is a foreground-weighted MSE on channel~0:
\begin{equation}
    \mathcal{L}_{\mathrm{heat}} = \frac{1}{HW}
    \sum_{i,j} w_{ij}\,\bigl(\hat{H}_{ij} - H_{ij}\bigr)^2 ,
    \qquad
    w_{ij} = 1 + \alpha\,\lvert H_{ij}\rvert ,
\end{equation}
where $\alpha = 25$ upweights pixels near blob centres (both positive
and negative) so that the loss from missing a blob is comparable to the
background loss.

The offset term is a masked smooth-$\ell_1$ loss on channels~1--2,
active only at end-point blob locations (where $H_{ij} < -0.05$):
\begin{equation}
    \mathcal{L}_{\mathrm{off}} = \frac{1}{N_{\mathrm{end}}}
    \sum_{i,j \in \mathcal{E}}
    \mathrm{SmoothL1}\!\Bigl(
        \frac{\hat{\mathbf{d}}_{ij} - \mathbf{d}_{ij}}
             {(W,\,H)}
    \Bigr) ,
\end{equation}
where $\hat{\mathbf{d}}_{ij}$ and $\mathbf{d}_{ij}$ are the predicted
and target offset vectors, $\mathcal{E}$ is the set of end-point pixels,
$N_{\mathrm{end}} = |\mathcal{E}|$, and the division by $(W, H)$
normalises the offsets to $[0, 1]$ so the loss scale is independent of
spatial resolution.  The offset weight is $\lambda_{\mathrm{off}} = 1$.

\section{Training Hyperparameters}
\label{app:hyperparams}

\Cref{tab:hyperparams} lists the optimiser and schedule settings
used to train the two models.  All training is performed on an NVIDIA
GH200 GPU.  \textsc{CSMClassifier} is trained from scratch on
experimental data; \textsc{ChargeLineNet} is pre-trained on synthetic
data and then fine-tuned on experimental data (see
\cref{subsec:finetuning}).

\begin{table}[ht]
\centering
\caption{Training hyperparameters for the two models.  The
\textsc{ChargeLineNet} column lists the fine-tuning values.}%
\label{tab:hyperparams}
\resizebox{\columnwidth}{!}{%
\begin{tabular}{lll}
\toprule
                          & \textsc{CSMClassifier} & \textsc{ChargeLineNet} \\
\midrule
Optimiser                 & AdamW            & AdamW \\
Learning rate             & $4\times10^{-3}$ & $10^{-4}$ \\
Head LR multiplier        & --               & $10\times$ \\
Weight decay              & $10^{-4}$        & $10^{-5}$ \\
LR schedule               & cosine           & cosine \\
Warm-up length            & short            & 5 epochs \\
Min.\ learning rate       & $10^{-7}$        & $10^{-6}$ \\
Gradient clip             & $1.0$            & $1.0$ \\
Batch size                & $32$             & $16$ \\
Max epochs                & $100$            & $100$ \\
Early-stop patience       & $30$             & $10$ \\
LR-plateau patience       & $7$              & -- \\
Bottleneck dropout        & $0.1$            & $0.2$ \\
Head dropout              & $0.1$ / $0.05$   & -- \\
Fine-tune freeze mode     & --               & encoder \\
\bottomrule
\end{tabular}%
}
\end{table}

\textsc{CSMClassifier} uses per-class positive weights to handle class
imbalance and the following on-the-fly augmentations: random vertical
and horizontal flips (50\,\% probability each) and resolution
augmentation (30\,\% probability, scale factor drawn uniformly from
$[0.8, 2.0]$).  \textsc{ChargeLineNet} fine-tuning additionally applies
domain-gap augmentation (\cref{app:domaingap}).

\section{Inference and Line Counting}
\label{app:inference}

\paragraph{Why the offset channels are needed.}
A signed heatmap alone detects start and end points independently but
does not encode which start belongs to which end
(\cref{subsec:lineheatmap}).  When lines are well separated, a simple
heuristic such as nearest-column matching or the Hungarian algorithm
pairs them reliably.  In practice, however, lines frequently overlap or
run close together: in the few-electron regime the inter-line spacing can
be comparable to the Gaussian blob radius, causing adjacent blobs to
merge.  When two end-point blobs overlap, a single connected component is
detected and only one line is counted; when a start blob from one line
overlaps an end blob from a neighbouring line, the two partially cancel,
reducing peak amplitude below the detection threshold.  Both failure
modes lead to systematic undercounting.  The offset channels resolve this
by encoding the pairing explicitly: each end-point blob carries a vector
pointing to its paired start, so the decoder recovers the correct pairing
even when blobs overlap spatially.  Because the offset is averaged over
the blob region (weighted by intensity), the estimate is robust to
per-pixel prediction noise and to partial blob merging: as long as the
blob centroid is correctly localised, the averaged offset still points to
the correct start.  This replaces the combinatorial matching step with a
per-blob vector readout whose cost is linear in the number of detected
blobs.

The offset is anchored at end points rather than start points because
the end point is the fading tip where the transition line trails off into
the background as the tunnel coupling is reduced along the exchange gate
axis.  This fading tip is visually unique and stands clearly apart from
the surrounding branching structure, making it spatially well-localised
and unambiguous in the heatmap.  Start points, by contrast, lie within
the branching structure where the transition begins, amid the
neighbouring parasitic-dot transitions, so their heatmap blobs are
harder to place precisely.  Anchoring the offset at the more reliable end
point ensures that the vector originates from a well-defined spatial
location, improving pairing accuracy.  This is borne out by the endpoint
localisation error on the held-out set (\cref{fig:per_occupancy_accuracy}c):
for matched lines, the median Euclidean distance between predicted and
ground-truth endpoints is $2.3$\,px for start points and $1.1$\,px for
end points (on $128 \times 128$ images), and the distributions are
skewed, with 90th-percentile errors of $12.0$\,px and $3.4$\,px
respectively.

\paragraph{Decoding procedure.}
At inference the three-channel \textsc{ChargeLineNet} output is decoded
into charge-transition lines as follows
(\cref{fig:chargelinenet_outputs} shows the input, predicted
heatmap, and decoded lines for three examples).  Negative peaks
(end-point blobs) are detected in channel~0 by thresholding and
connected-component labelling.  Rather than a fixed threshold, we use a
scale-invariant adaptive policy: the detection threshold is set to the
larger of a noise-anchored level (the heatmap median plus
$5\sigma_{\mathrm{noise}}$) and a fixed fraction ($0.25$) of the signal
ceiling.  A detected line is retained only if either of its endpoints
peaks above a stronger hysteresis confirmation threshold ($0.5$ of the
signal ceiling), which rejects noise on near-blank images while keeping
faint-ended lines that have a confident start.  On densely populated
maps this confirmation threshold is relaxed relative to the image's own
median blob peak to avoid undercounting.

For each end-point blob, the offset vector (channels~1--2) is averaged
over the blob region, weighted by blob intensity, to obtain a robust
estimate of the end-to-start displacement.  The predicted start position
is computed as the blob centroid plus the averaged offset.  This
offset-predicted start is then used as the seed for a snapping step: a
corridor is searched along the offset direction and the seed is snapped
onto the nearest detected positive peak (start blob), scored by a
combination of distance along and perpendicular to the offset ray and
angular alignment with it.  This fusion is necessary because neither
ingredient suffices alone: the offset vector resolves the pairing (which
end belongs to which start) and supplies a search direction, but as a
continuous regression output its predicted start location carries an
error of several pixels; the independently detected start blob localises
the start precisely but carries no information about which end it belongs
to.  Snapping the offset-predicted start onto the nearest detected start
blob combines the two, so that the vector resolves the association while
the blob corrects the regression error and pins the start to a sharply
localised position.  Without the blob the start would inherit the
regression error, and without the vector the pairing would require a
separate combinatorial matching step that fails when blobs merge.  Duplicate line pairs that collapse to the same start--end
coordinates after snapping are removed.  Finally, lines shorter than a
minimum length or whose angle from vertical exceeds a configurable
threshold are discarded.  The electron count is the number of surviving
line pairs.

\section{High-Occupancy Sampling and Extension}
\label{app:occupancy}

The per-occupancy breakdown (\cref{fig:per_occupancy_accuracy}a) shows
that accuracy does not degrade as the line count grows, but the highest
bins are estimated from few images, so their per-bin accuracies should be
read as indicative rather than precise.  This under-representation is not a labelling choice but a
consequence of data collection: experimental CSMs are costly to acquire,
and obtaining images with a prescribed number of transition lines is not
straightforward, since the electron occupancy is set by device tuning
that is itself the difficult step the pipeline is meant to assist.  The
aggregate figure is therefore dominated by the low-occupancy regime,
which is both the regime of practical interest and the best represented
in the data, but the per-occupancy breakdown confirms that this weighting
does not mask a hidden weakness at high counts.

We expect the training pipeline to extend gracefully to still higher
occupancies than are sampled here: the resolution augmentation already
exposes the model to a range of line densities, and the synthetic
pre-training plus transfer-learning route makes it inexpensive to extend
coverage of the high-occupancy regime once such data becomes available,
at which point a wider, landscape-format scan window would help resolve
transitions that begin to overcrowd the fixed square $128\times128$
input.

\section{Domain Gap and Freeze-Mode Selection}
\label{app:domaingap}

Experimental images differ from synthetic ones in several ways \cite{Zwolak2024-tf}, which
together constitute the domain gap that fine-tuning
(\cref{subsec:finetuning}) is designed to close:
\begin{itemize}
    \item \textbf{Global contrast and brightness.}  These vary between
          devices and with the tuning of the charge sensor (SET),
          producing images whose intensity distributions differ from the
          simulator output.
    \item \textbf{Background structure.}  Electrical noise, such as
          50\,Hz noise, a slope in the background due to capacitive
          coupling to the SET and/or malfunctioning of the feedback, and
          telegraph noise produce spatially-varying backgrounds that are
          absent from synthetic data.
    \item \textbf{Scan artefacts.}  Voltage-sweeping patterns can
          produce sensor artefacts and charge hysteresis that the
          simulator does not generate.
    \item \textbf{Device-specific features.}  Parasitic charge
          transitions from nearby dots, cross-talk between gate lines,
          electron tunnelling rates, and other non-ideal device
          behaviour are approximated by the simulator, but their exact,
          device-specific appearance varies from device to device and
          cannot be reproduced exactly.
\end{itemize}

The choice of freeze mode during fine-tuning trades off adaptation
against stability.  Freezing more layers reduces the risk of catastrophic
forgetting but limits how much the model can adjust to the new data
distribution, while freezing fewer layers allows greater adaptation but
requires more labelled data to avoid overfitting.  Freezing the encoder
(as used throughout this work) holds the feature-extraction backbone
fixed while the task-specific decoder and output head are retrained,
which provides a good balance for the $6{,}452$-image fine-tuning set:
it adapts the output stages to the experimental domain while retaining
the structural features learned during synthetic pre-training.  This
leaves $258{,}659$ of \textsc{ChargeLineNet}'s $935{,}283$ parameters
trainable during fine-tuning (about $28\,\%$).

On-the-fly domain-gap augmentation is applied during fine-tuning to
bridge the categories above.  Random background noise, scan-line stripes,
dither patterns, and parasitic transition lines are added to each input
image at every training epoch, so the model sees a wider range of
artefact combinations than the small hand-labelled dataset alone
provides.  Augmentation is applied only to the input images; heatmap and
coordinate targets are left unchanged.

\section{Synthetic Pre-training Data}
\label{app:synthetic}

\begin{figure*}[t]
\centering
\includegraphics[width=0.72\textwidth]{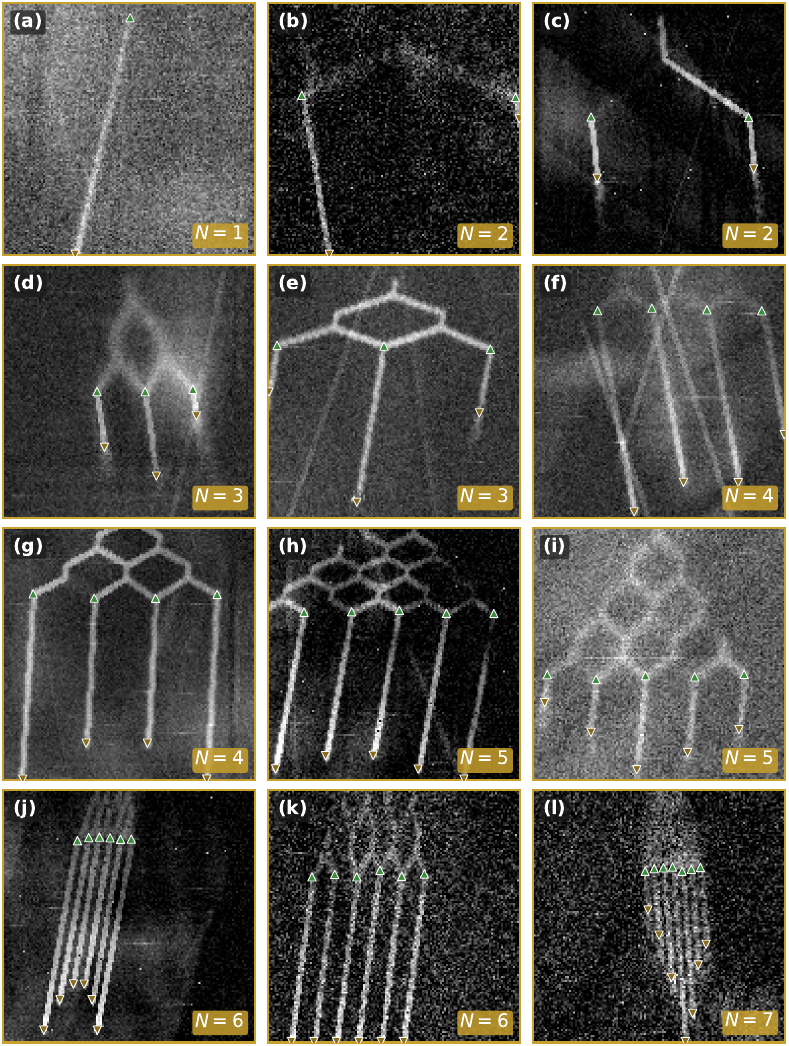}
\caption{Representative synthetic CSM images produced by the
parameterised image simulator and used to pre-train
\textsc{ChargeLineNet} (twelve examples, spanning line counts $N=1$ to
$N=7$).  Each panel is a $128\times128$ generated image; the gold label
gives the ground-truth number of plunger-gate transition lines $N$.
Green upward markers ($\blacktriangle$) and bronze downward markers
($\blacktriangledown$) show the point-based supervision targets, namely the
sharp start point and the blur-onset end point of each final line, which
are the only annotations \textsc{ChargeLineNet} is trained on.  The
generator reproduces the salient features of experimental isolated-mode
CSMs: near-vertical final transition lines that fade toward their tips,
the honeycomb-like branching structure of higher-level transitions that
must be excluded from the count, variable line spacing, contrast, and
lever-arm tilt, and a broad range of artefacts (background gradients and
vignetting, scan-line and column-ripple pickup, trap and coupling
stripes, diffuse blobs, and measurement noise).}%
\label{fig:synthetic_examples}
\end{figure*}

The \textsc{ChargeLineNet} pre-training corpus (${>}10^5$ images) is
produced by a lightweight parameterised \emph{image} simulator rather
than a physics-based device simulation: it draws the transition lines as
geometric segments and renders them directly, then adds background
structure and measurement noise, so images are generated at negligible
cost and with exact ground-truth start/end coordinates.  Each image
samples a random line count, spacing, contrast, lever-arm tilt, line
fading, and a suite of noise and artefact models
(\cref{subsec:finetuning}).  Representative synthetic images,
together with the point-based supervision targets they provide, are shown
in \cref{fig:synthetic_examples}.

% ---- Synthetic pre-training example images ----

\section{Classical Baseline Comparison}
\label{app:classical}

We benchmarked a set of classical line counters against
\textsc{ChargeLineNet} on the same 1{,}131-image held-out validation set,
using the same preprocessing, Hungarian line matching, and count metrics
(\cref{subsec:results_chargelinenet_perf}).  Each detector, together with
its classical image-preprocessing stage, exposes a set of tunable
hyperparameters, 16 in total across detector and preprocessing (for
example the Canny edge thresholds, the Hough accumulator and angular-gate
settings, and the bilateral-denoise strength).  These were optimised by
grid search on a random $10\,\%$ subsample drawn from all images of the 16
training devices, held strictly separate from the 16 held-out validation
devices, which the classical pipeline never sees, so the comparison is on
the same cross-device footing as \textsc{ChargeLineNet}.

The classical baseline's accuracy is not sensitive to how
these knobs are set.  Repeating the tuning on five independent random
$10\,\%$ subsamples of the training images yields five different optimal knob settings
but essentially the same held-out accuracy (varying by well under one
percentage point across the five runs).  The gap to \textsc{ChargeLineNet}
therefore reflects a genuine limitation of the classical strategies rather
than an unlucky choice of hyperparameters.

\paragraph{Detectors.}
We evaluated three detectors spanning the classical strategies named in the
introduction:
\begin{itemize}
    \item \textbf{Hough.}  A probabilistic Hough transform on a Canny edge
          map, keeping only near-vertical segments (angular gating) and
          clustering collinear segments in the same image column into a
          single line.
    \item \textbf{Profile.}  A vertical-matched-filter column projection: the
          image is smoothed strongly along the line direction (to
          raise the line-to-noise ratio), a second-derivative-of-Gaussian
          matched filter is applied across the orthogonal direction, and the
          rectified response is integrated over all rows into a single
          one-dimensional column profile whose peaks are the detected lines.
    \item \textbf{Row-peaks.}  A per-row peak counter that exploits the
          physical structure of the branching: genuine plunger-gate lines
          span the full scan height whereas parasitic-dot branching appears
          only in part of the image, so each horizontal row is scanned for
          peaks and the (robust) minimum peak count across rows in the lower
          band of the image is taken as the line count.
\end{itemize}

\paragraph{Classical preprocessing.}
We tuned a classical preprocessing pipeline (background flattening by
large-kernel subtraction, edge-preserving denoising, and contrast-limited
adaptive histogram equalisation, CLAHE) jointly with each detector.  The
clear best choice was edge-preserving bilateral denoising, which
suppresses the background noise texture that would otherwise spawn spurious
edges while preserving the sharp transition edges; it improved the Hough
exact-count accuracy from $50\,\%$ to $61\,\%$.  CLAHE and
background subtraction, by contrast, degraded performance, because they amplify
background structure into additional false lines.  This suggests that the
difficulty is not one of global contrast but of distinguishing genuine
transitions from spurious ones, the selectivity that a learned model
provides.

\begin{figure*}[t]
\centering
\includegraphics[width=0.98\textwidth]{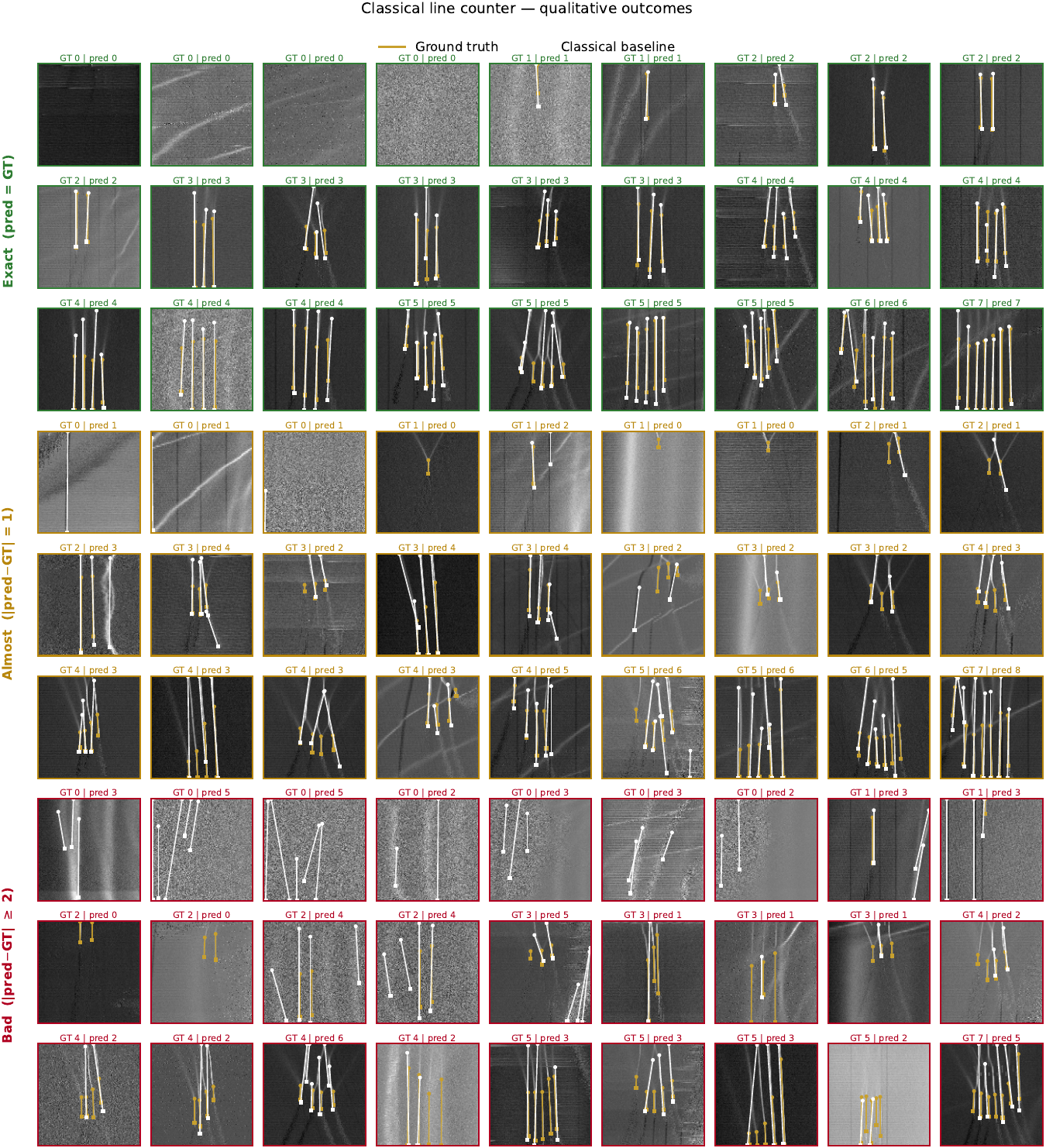}
\caption{Qualitative outcomes of the classical line counter (probabilistic
Hough transform with the bilateral-denoise preprocessing of
\cref{app:classical}, the strongest of the classical detectors on
exact-count accuracy) on 81 held-out CSMs, grouped into three bands:
\emph{exact} (predicted count equals ground truth; green), \emph{almost}
(off by one; gold), and \emph{bad} (off by two or more; red).  Ground-truth
plunger-gate lines are drawn in gold and classical detections in white
(start = circle, end = square), overlaid on the preprocessed image; each
panel title gives the ground-truth and predicted line counts.  The exact
band shows the regime where a classical counter genuinely succeeds --
correctly returning zero on featureless maps and counting well-separated,
low-to-moderate-occupancy lines.  The bad band shows the systematic failure
modes that motivate a learned detector: spurious counts on scan-line and
background-gradient artefacts (e.g.\ GT~0 counted as several lines),
over-counting of honeycomb branching from parasitic dots, and collapse in
the dense or faint high-occupancy regime.}%
\label{fig:classical_baseline_grid}
\end{figure*}

\paragraph{Results.}
\Cref{tab:classical_baseline} reports the best configuration of each
detector against \textsc{ChargeLineNet}.  Even after this hyperparameter and
preprocessing optimisation, the strongest classical detector reaches
only $61\,\%$ exact-count accuracy, roughly $34$ percentage points below
\textsc{ChargeLineNet}.  This is, coincidentally, almost exactly the
$61.4\,\%$ reached by the synthetic \textsc{ChargeLineNet} before
fine-tuning (\cref{tab:linecount}); the two very different baselines, one
learned but un-adapted and one hand-engineered, happen to sit the same
${\sim}34$ points below the fine-tuned model.  The gap is the direct consequence of the miscounting
mechanism described in \cref{sec:introduction}: lacking any learned
notion of which transitions are genuine, these detectors respond to branching
and artefacts as readily as to plunger-gate lines.  The three detectors also
trade off precision against recall in incompatible ways, and each is limited
by a distinct failure mode:
\begin{itemize}
    \item \textbf{Hough} attains the highest exact-count accuracy but the
          lowest per-line precision ($0.42$).  Its errors are roughly
          balanced between over- and under-counting ($19\,\%$ and $20\,\%$ of
          images), which is why it scores well on aggregate count while still
          matching individual lines poorly: over- and under-detections
          partially cancel in the count even when the detected lines are
          misplaced.  Its single largest weakness is spurious response to
          artefacts on featureless maps: on the $161$ ground-truth-empty
          (zero-line) images it wrongly reports one or more lines
          $42\,\%$ of the time (\cref{fig:classical_baseline_grid}, bad
          band, GT~$0$ panels, where scan-line ripple and background gradients
          are counted as vertical transitions).
    \item \textbf{Profile} localises lines the best (highest precision $0.70$
          and $F_1$ $0.65$): its vertical matched filter accumulates evidence
          down each column, so the lines it does report are well placed.  But
          for the same reason it cannot suppress branching, because a branch
          adds to the column response exactly like a genuine line, so it
          over-counts on branched maps, and its strong along-line smoothing
          causes it to merge closely-spaced transitions, collapsing its
          exact-count accuracy in the high-occupancy regime (zero exact
          matches at seven lines).
    \item \textbf{Row-peaks}, built specifically to reject branching by taking
          the minimum peak count across rows, is the most brittle: the
          minimum operator is dominated by whichever row is noisiest or where
          a faint line locally fades, so it under-counts most often ($F_1$
          $0.43$, exact $35\,\%$) and never becomes competitive.
\end{itemize}
Across the shared examples in \cref{fig:classical_baseline_grid}, the
off-by-one (\emph{almost}) band is split fairly evenly between over- and
under-counts, while the \emph{bad} band is dominated by two mechanisms:
artefact hallucination on low-occupancy maps (the GT~$0$--$2$ over-counts) and
the merging or missing of faint, closely-spaced lines at moderate occupancy
(the GT~$3$--$5$ under-counts).  The classical counters do
not fail preferentially at high occupancy.  Hough is in fact most
accurate there ($65$--$76\,\%$ exact at five to seven lines), because those
maps tend to be cleaner and better-separated, so the errors reflect
artefact and branching confusion rather than raw line density.  No single
classical strategy achieves both accurate counting and clean localisation,
whereas \textsc{ChargeLineNet} does both at once.
\Cref{fig:classical_baseline_grid} shows representative exact,
off-by-one, and badly-miscounted outcomes for the strongest (Hough) detector.

\begin{table}[h]
\centering
\caption{Classical line counters versus \textsc{ChargeLineNet}
on the 1{,}131-image held-out set.  Each classical detector uses
hyperparameters and bilateral-denoise preprocessing optimised on a random
$10\,\%$ subsample of all images from the 16 training devices (not the
held-out set);
\textsc{ChargeLineNet} numbers are the fine-tuned model from
\cref{subsec:results_chargelinenet_perf}.  ``Exact'' and ``$\pm1$''
are line-count accuracies; precision, recall, and $F_1$ are per-line
detection metrics at a $15$-px matching tolerance.}%
\label{tab:classical_baseline}
\begin{tabular}{lccccc}
\toprule
Method              & Exact (\%) & $\pm1$ (\%) & Prec. & Rec. & $F_1$ \\
\midrule
Hough               & 61 & 89 & 0.42 & 0.42 & 0.42 \\
Profile             & 52 & 75 & 0.70 & 0.61 & 0.65 \\
Row-peaks           & 35 & 63 & 0.42 & 0.44 & 0.43 \\
\midrule
\textsc{ChargeLineNet} & \textbf{95.3} & \textbf{98.6} & \textbf{0.93} & \textbf{0.92} & \textbf{0.92} \\
\bottomrule
\end{tabular}
\end{table}

% ---- Classical baseline: qualitative outcomes grid ----

% Flush all pending floats (the synthetic-examples figure) before the
% bibliography so no figure lands among the reference pages.
\clearpage

% =====================================================
\bibliography{references}
  
\end{document}